\documentclass[useAMS,usenatbib]{mn2e}

\usepackage{amsmath}
\usepackage{comment}
\usepackage{graphicx}
\usepackage{rotating}
\usepackage{color}
\usepackage{amsfonts}
\usepackage{epsfig}
\newcommand{\beq}{\begin{equation}}
\newcommand{\enq}{\end{equation}}
\newcommand{\beqa}{\begin{eqnarray}}
\newcommand{\enqa}{\end{eqnarray}}
\newcommand{\beit}{\begin{itemize}}
\newcommand{\enit}{\end{itemize}}
\newcommand{\bem}{\begin{pmatrix}}
\newcommand{\enm}{\end{pmatrix}}

\newcommand{\vectheta}{\bm{\theta}}

\newcommand{\vecq}{\mathbf{q}}

\newcommand{\lat}{\left\langle}
\newcommand{\rat}{\right\rangle}
\newcommand{\av}[1]{\lat #1 \rat}
\newcommand{\Var}[1]{\textrm{Var}\left(#1\right)}

\newcommand{\lp}{\left (}
\newcommand{\rp}{\right )}

\newcommand{\bes}{\begin{sideways}}
\newcommand{\ees}{\end{sideways}}

\renewcommand{\bbeta}{\boldsymbol {\eta}}
\renewcommand{\bmu}{\boldsymbol {\mu}}

\newcommand{\vecN}{\mathbf N}
\newcommand{\vecrho}{\boldsymbol \rho}
\newcommand{\vecA}{\mathbf A}

\renewcommand{\vectheta}{\boldsymbol \theta}
\newcommand{\vecd}{\mathbf d}

\title[Unveiling the cosmological information beyond linear
scales]{Unveiling the cosmological information beyond linear scales: forecasts for sufficient statistics}
\author[Wolk, Carron and Szapudi]{M. Wolk\thanks{E-mail:
wolk@ifa.hawaii.edu}, J. Carron and I. Szapudi  \\
Institute for Astronomy, University of Hawaii, 2680 Woodlawn Drive, Honolulu, HI, 96822}
\begin{document}

\date{\today}

\pagerange{\pageref{firstpage}--\pageref{lastpage}} \pubyear{2014}

\maketitle

\label{firstpage}
\begin{abstract}
Beyond the linear regime, Fourier modes of cosmological random fields become
correlated, and the power spectrum of density fluctuations contains only a fraction of the 
available cosmological information. To unveil this formerly hidden information, the $A^*$
non-linear transform was introduced; it is optimized both for the nonlinearities induced by gravity and observational noise. Quantifying the resulting increase of our knowledge of cosmological
parameters, we forecast the constraints from the angular power
spectrum and that of $A^{\ast}$ from $\ell \sim 200$ to
$3000$ for upcoming galaxy surveys such as: the Wide-Field Infrared Survey Telescope
(WFIRST), the Large Synoptic Survey Telescope (LSST), Euclid, the
Hyper Suprime-Cam (HSC) and the Dark Energy Survey (DES). We find
that at low redshifts this new data analysis strategy can double
the extracted information, effectively doubling the survey area.
To test the accuracy of our forecasting and the power of our data analysis methods, 
we apply the $A^*$ transformation to the latest release of the
Canada-France-Hawaii-Telescope Legacy Survey (CFHTLS) Wide. 
While this data set is too sparse to
allow for more than modest gains ($\sim 1.1-1.2$), the realized gain from
our method is in excellent agreement with our forecast, 
thus verifying the robustness of our analysis and prediction pipelines.
\end{abstract}


\section{Introduction}

Within the successful inflationary paradigm of cosmology, the small initial density fluctuations obey Gaussian statistics. This fact makes power spectra particularly powerful summary statistics: at early times, the amplitude of each wave number carries independent information, and the variance calculated from all of them contains all available information: the random phases carry no cosmological significance. In this case, an hypothetical ideal observation of the spectrum unlocks an amount of information  proportional to the number of resolved Fourier modes. For this reason, the power spectrum is among the most widely used statistic to characterize the large scale structures in the
Universe. From it, a wealth of information is to be gained constraining cosmological
models as it was successfully shown using large galaxy surveys such as the 2-degree field galaxy survey
\citep[e.g.,][]{Coleetal05} and the Sloan Digital Sky Survey
\citep[e.g.,][]{Tegmarketal04}. Moreover, the next decades will see the advent of large wide-field surveys that are designed to measure in exquisite detail the two-point statistics
of the matter field through weak-lensing or clustering, 
ultimately targeting fundamental questions such as the nature of
dark energy or neutrino masses.
\newline
\indent
Unfortunately, there are significant obstacles to clear before the new generation of surveys can achieve their worthwhile goals. In particular,  the observational noise and the correlation of Fourier modes developing from non-linear gravitational growth decrease the amount of information accessible to the power spectrum.  In the mildly non-linear
regime the information saturates at a finite plateau instead of growing sharply with the cube (or square, in the case of projected density) of the maximal resolved wavenumber. At first sight this
leads to an impressive mismatch with naive Gaussian expectations
\citep{Rimesetal05,Rimesetal06,Neyrincketal06,Neyrincketal07}. It is
now well understood that this is mostly due to a finite volume effect \citep{Szapudi96}
built out of two main components: first the
``beat coupling'' correlating small scales to the survey scale and then the arguably large background mode variance even for large surveys \citep{Rimesetal05, Rimesetal06,
  dePutteretal12, Takadaetal13}. In particular, in the case of the noise free dark matter field, inference on traditional cosmological parameters generically suffers substantially from the necessary calibration or reconstruction of the local density \citep{Carronetal14b,LiEtal14}.
\newline
\indent
Since all the upcoming cosmological surveys are demanding in resources, it is worth
investigating alternative analysis
strategies than the power spectrum, with the aim of being more efficient. Mainstream methods include most notably higher-order $N$-point statistics
\citep[e.g.,][]{Peebles1980,Szapudi2009}. The analysis of higher order
statistics are however often difficult due to a steep combinatorial complexity, and furthermore gravity drives the matter field towards a regime where they should not be expected to capture  information efficiently \citep[and references therein]{CarronNeyrinck2012}. Introduced specifically with efficiency in mind are non-linear transformations, such as the logarithmic
mapping \citep{Neyrincketal09} or variants thereof \citep{Seoetal11,Joachimietal11}. 
While these transforms were originally phenomenologically motivated, \cite{Carronetal13} demonstrated how to construct explicitly
transforms of the field that capture by design most of the
available information, resulting in approximate ``sufficient
statistics''. Then, \cite{Carronetal14a}, taking further into account
the discreteness effects in galaxy surveys introduced the $A^*(N)$
non-linear transformation as the optimal observable to extract the
information content. The latter is the analog of the logarithmic dark matter field $\delta$ transform $A = \ln(1 + \delta)$ for galaxy count maps $N$ to which it reduces for large sampling rates.
\newline
\indent
Non-linear transformations, and more specifically the optimal $A^*$ transformation, take into account discreteness effects and the non-Gaussianity of the field to improve the statistical power of the spectrum. However, it remains somewhat unclear to this day how much overall improvement over standard methods one can expect from this approach. The principal aim of this paper
is to estimate the constraining power of the $A^*$ angular power spectrum compared to that of the angular galaxy power spectrum on cosmological parameters in realistic current and future projected surveys.
\newline
\indent
In order to address this question, we need to capture in a satisfying manner all the effects
discussed above. Throughout this work, we adopt a halo occupation
distribution model \citep[hereafter HOD, ][]{Seljak00, Scoccimarro01,
  Kravtsovetal04, Zhengetal05} to describe the galaxy
clustering. The cosmic variance is taken into account using the assumption of lognormal field statistics for a $2d$ projected field and the projected galaxy counts are
described by multinomial sampling of that field. 
\cite{Carronetal14c} have demonstrated that these prescriptions reproduce
accurately the statistical properties of the galaxy field.
\newline
\indent 
We proceed as follows. Section~\ref{sec:model} presents the modeling that
enters our predictions. Described in Section~\ref{sec:maps} is the
fast projected galaxy counts map simulation pipeline that we use,
tested on CFHTLS data for accuracy. 
With this tool we forecast the expected gain in information of the $A^*$ transform for different surveys and parameters, as presented in Section~\ref{sec:est}. Our predictions are also compared to actual measurements from the CFHTLS data. We summarize and conclude with a discussion in Section~\ref{sec:discuss}. Two appendices collect additional technical details of the methodology.

\section{Modeling}
\label{sec:model}

This section reviews the different ingredients that enter our analysis of the data and forecasts later in section \ref{sec:maps}. The statistical aspects of the model are discussed in \ref{subsec:stat}. The model takes as necessary input the galaxy two-point correlation function, for which we use the HOD parametrization discussed in \ref{subsec:HOD}. Finally, our fiducial values for the cosmological and HOD parameters are discussed in \ref{sec:fid}.
\subsection{Statistical modeling}
\label{subsec:stat}
Let $\vecN = (N_1,\cdots,N_{d})$ be a map of galaxy counts in $d$
cells containing respectively $N_{1}, ..., N_{d}$ objects. We model
the map as a discrete sampling of an underlying continuous galaxy field $\vecrho_g = (\rho_{g,1},\cdots,\rho_{g,d})$. This field is chosen to obey lognormal field statistics, i.e. the map $\vecA = \ln \vecrho_g$ is Gaussian. The $d \times d$ covariance matrix of the Gaussian field is related to that of the galaxy field through
\beq
\omega_{A,ij} = \ln(1 + \omega_{\delta_{g,ij}}),
 \label{eq:xiAxid}
\enq
where $\omega_{\delta_g}$ is the galaxy two-point function discussed in section \ref{subsec:HOD}, further filtered as described in the Appendix of \cite{Carronetal14c} to account for the slight anisotropy induced by the square cells.
\newline
\indent
To complete the statistical description of the counts, we need discrete sampling of the
underlying continuous field. To do so, there at least are two natural choices:
Poisson sampling, for which the number of galaxies varies from one map to
the other, and multinomial sampling for which the total number of
objects is the same in each map. In this study we use the latter. This
choice has the main virtue of simplifying the interpretation of the
results as there is no need to introduce a Poisson sampling intensity
parameter $\langle N \rangle = \bar{N}$, and to marginalize over it
in the end. Arguably, one may worry about the difference in cosmic variance and how
well it is taken into account compared to a Poisson sampling. 
As discussed in Appendix~\ref{app:error}, in practice both sampling methods lead to identical results when performed consistently.
\newline
\indent
Let $f_i$ be the unmasked fraction of cell $i$.   We set the multinomial sampling probability in cell $i$ to be proportional to that fraction times the galaxy field ${\rho_{g,i}}$ behind it. Explicitly, the probability for the count map $\vecN$ in the presence of the galaxy field may be written as
\beq \label{sampling}
P(\vecN|\vecrho) = \lp \frac{N_\textrm{tot} !}{N_1!\cdots N_d!} \rp {\prod_{i = 1}^d \lp \frac{f_i\rho_i}{\sum_j f_j \rho_j}\rp^{N_i}},
\enq
with $N_\textrm{tot} = \sum_i N_i$ is the total
number of galaxies in the map. 

\subsection{Analytical modeling}
\label{subsec:HOD}

The key quantity in the above section is the galaxy angular two-point
correlation function $\omega(\theta)$. We found that in
order to correctly reproduce the behavior of the galaxy field on small scales
we need to take into account how galaxies are distributed within the dark
matter haloes. To do so, we use the ``halo
model'' \citep{Scoccimarro01, Maetal00, Peacock00,
  Coorayetal02} which states that the galaxy-galaxy correlation
function can be written as a sum of two contributions:
\beq
\xi(r) = \xi_{1h}(r) + \xi_{2h}(r).
\label{eq:2pcf}
\enq
The first term, called the one-halo term, comes from pairs of galaxies that reside within the same dark
matter halo and depends on the number of galaxy pairs per halo
$\langle N (N-1) \rangle$. The second term, called the two-halo term,
is due to pairs of galaxies that reside in two separate dark matter
haloes and depends on the number of galaxies per halo $\langle N \rangle$.
To calculate the galaxy clustering, we need to describe how galaxies populate dark matter
haloes and to do so, we closely follow \cite{Zhengetal07} which describes
$N(M)$, the number of galaxies in a halo of given mass $M$, as a sum of two
terms: one coming from the central galaxy in the halo $N_{c}(M)$ and
the other coming from the satellites $N_{s}(M)$. Thus $N(M)$ can be
expressed as:
\begin{equation}
N(M)=N_{c}(M) \times [1+N_{s}(M)]
\end{equation}
where
\begin{equation}
N_{c}(M)=\frac{1}{2}\Big[1+\textrm{erf}\Big(\frac{\log M-\log
  M_{min}}{\sigma_{\log M}}\Big)\Big],
\end{equation}
and
\begin{equation}
N_{s}(M)=\Big(\frac{M-M_{0}}{M_{1}}\Big)^{\alpha}.
\end{equation}
Our model has five adjustable parameters: $M_{\rm min}$, $M_{1}$,
$M_{0}$, $\alpha$ and $\sigma_{\log M}$. 
For the halo mass function, we use the
prescription from \cite{Sethetal99}.
Furthermore, we describe the halo density profile using a
Navarro-Frenk-White (NFW) profile \citep{Navarroetal97} and we assume
that haloes are biased tracers of the matter distribution using for
the halo bias, $b_{h}(M,z)$, the parametrization from \cite{Tinkeretal05} calibrated
on simulations. More details can be found in \cite{Couponetal12}.
The knowledge of the cosmology and of the above five HOD parameters allow us to estimate the
two-point correlation via Equation~\ref{eq:2pcf} and thus its
projected counterpart $\omega$.

\subsection{Fiducial HOD parameters and cosmology}
\label{sec:fid}

Our fiducial model for the HOD parameters is based on galaxy observations using the seventh and final version of the
Canada-France-Hawaii-Telescope Legacy survey
(CFHTLS)\footnote{http://www.cfht.hawaii.edu/Science/CFHLS/} and the sample selection of \cite{Wolk13}.


For the purpose of that paper, we restrict our study to the biggest
field, W1, which has the highest statistics.  The W1 field is
approximately a square of $L = 7.46$ degrees on the side, that we
divide into $128^{2}$ square cells. Doing so, we can probe the galaxy
angular power spectrum in the multipole range $240 \le k \le 3100$. 
We consider four redshift bins: $0.2 < z < 0.4$, $0.4 < z < 0.6$, $0.6 < z < 0.8$ and $0.8 < z < 1.0$. A large bin width ($\Delta_{z} = 0.2$) ensures a low bin-to-bin contamination. 

The cosmological and HOD parameters used in \cite{Wolk13} are summarized in Tables~\ref{table:cosmo} and \ref{table:HOD}. These best-fit HOD parameters were derived fitting the angular two-point
correlation using the Population Monte
Carlo (PMC) technique as implemented in the
\texttt{CosmoPMC}\footnote{\texttt{http://cosmopmc.info}} package.  
This fiducial model fixes the $A^{\ast}$-mapping parameters, presented
in Table~\ref{table:mapping} and moreover it means that, by
construction, all the CFHTLS simulated maps have the two-point
statistics determined by this fiducial model.

\begin{table}
\centering
\caption{Cosmological model parameters. The parameters marked with a dagger are kept fixed when analysing the model for the fiducial case.}
\begin{tabular}{cccccccc}
\hline
\hline
$\Omega_{m}$    &  $\Omega_{K}^{\dagger} $&$\Omega_{b}^{\dagger}$
&$w_{0}$ &$w_{a}^{\dagger}$ & $h^{\dagger}$
&$n_{s}^{\dagger}$ & $\sigma_{8}$  \\
0.27 & 0.0$^{\dagger}$ & 0.045$^{\dagger}$ & -1.0 &0.0$^{\dagger}$ &0.70$^{\dagger}$ &  0.96$^{\dagger}$ &  0.80 \\
\hline
\hline
\end{tabular}
\label{table:cosmo}
\end{table}

\begin{table*}
\centering
\caption{HOD model parameters obtained by fitting the measured angular two-point
  correlation function on the CFHTLS data.}
\begin{tabular}{cccccc}
\hline
\hline
Redshift bin    & $\log M_{min}$ & $\log M_{1}$ & $\log M_{0}$ &
$\sigma_{\log M}$ & $\alpha$ \\
\hline
$0.2<z<0.4$ & 12.13$^{+0.11}_{-0.16}$   &
13.24$^{+0.06}_{-0.05}$ & 9.67$^{+1.15}_{-1.05}$ &
0.75$^{+0.18}_{-0.29}$ & 1.15$^{+0.03}_{-0.03}$\\

$0.4<z<0.6$ &  12.02$^{+0.09}_{-0.17}$ &
13.00$^{+0.06}_{-0.06}$&11.93$^{+0.15}_{-0.20}$&
0.79$^{+0.14}_{-0.36}$&0.98$^{+0.04}_{-0.05}$ \\

$0.6<z<0.8$ &
12.04$^{+0.05}_{-0.06}$&12.95$^{+0.05}_{-0.05}$&11.80$^{+0.15}_{0.20}$&0.94$^{+0.04}_{-0.04}$&0.99$^{+0.05}_{-0.05}$\\

$0.8<z<1.0$ &
12.21$^{+0.05}_{-0.05}$&13.07$^{+0.08}_{-0.10}$&12.42$^{+0.14}_{-0.15}$&0.96$^{+0.02}_{-0.07}$&0.79$^{+0.12}_{-0.13}$
\\
\hline
\hline
\end{tabular}
\label{table:HOD}
\end{table*}

\begin{table}
\centering
\caption{$A^{\ast}$-mapping parameters for the CFHTLS. $\bar{N}$ is an
  estimation of the sampling rate and $\sigma^{2}_{A}$ is the variance
of the $A$ field related to the variance of the galaxy
field by $\sigma^2_A = \ln (1 + \sigma^2_{\delta_g})$.}
\begin{tabular}{ccccc}
\hline
\hline
Redshift bin    & $\sigma^{2}_{A}$ & $\bar{N}$ \\
\hline
$0.2<z<0.4$ & 0.274   &   2.086 \\
$0.4<z<0.6$ &  0.172   &    5.861\\
$0.6<z<0.8$ & 0.135     &  9.684\\
$0.8<z<1.0$ &  0.120  &    8.836\\
\hline
\hline
\end{tabular}
\label{table:mapping}
\end{table}

\subsection{Surveys specificities}
\label{subsec:spec}

As a non-exhaustive but however representative ensemble of upcoming
galaxy surveys, we choose to consider configurations close to the ones
expected for the Hyper
Suprime-Cam\footnote{http://www.naoj.org/Projects/HSC/} (HSC),
Euclid\footnote{http://sci.esa.int/euclid/}, the Large Synoptic Survey
Telescope\footnote{http://www.lsst.org/lsst/} (LSST), the Dark Energy
Survey\footnote{http://www.darkenergysurvey.org/} (DES) and the
Wide-Field Infrared Survey
Telescope\footnote{http://wfirst.gsfc.nasa.gov/} (WFIRST).

\begin{figure}
  \begin{center}
    \includegraphics[width=9cm]{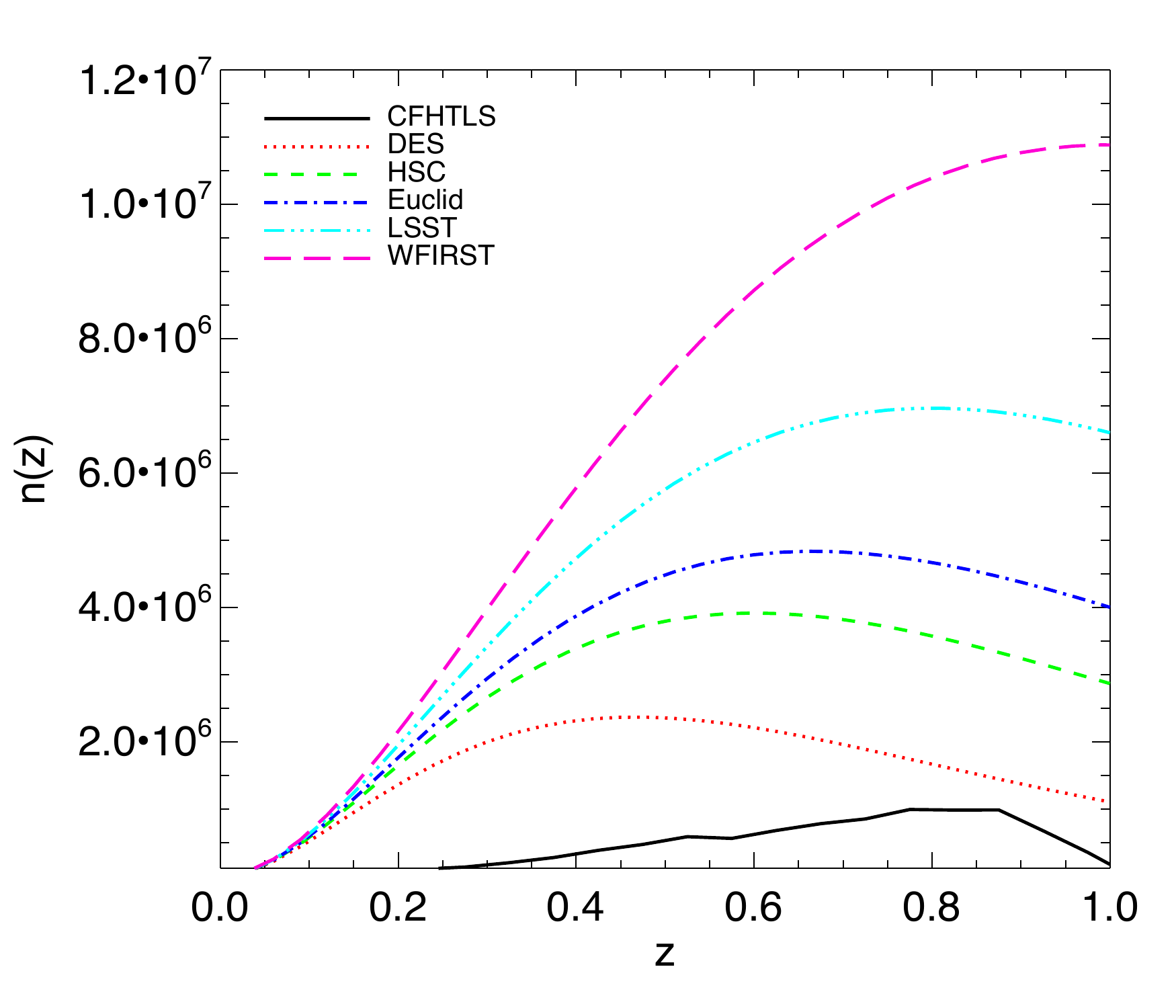}
  \caption{\label{figure:dndz} Photometric redshift distributions used in the forecasts
    for the different upcoming surveys. The latter are given by a
    simple 1-parameter
    analytic function described in Equation~\ref{eq:nofz}. In the case
  of the CFHTLS-W1 field the redshift distribution is estimated from the data.}
\end{center}
\end{figure}

Following \cite{Takadaetal09} we model the redshift distribution of
the objects through the following one-parameter functional form:


\beq
\frac{dn(z)}{dz} = n_{0} \times 4z^{2} \exp \Big(-\frac{z}{z_{0}} \Big)
\label{eq:nofz}
\enq
where the normalization is fixed to $n_{0} \simeq 100$
arcmin$^{-2}$ and $z_{0}$ is related to the mean redshift, $z_{m}$, by $z_{0} =
z_{m}/3$. The only free parameter is $z_{m}$ and
Table~\ref{table:surveys_carac} shows the values that we pick for each survey. Also shown are the angular number
density of galaxies $\bar{n}_{g} \equiv \int_{0}^{+\infty} dz \,
dn(z)/dz$. Figure~\ref{figure:dndz} shows the distribution of the number density of
objects over the redshift range $z \in [0, 1.0]$. In the case of the CFHTLS-W1 field, we use directly the measured redshift distribution. We build our samples
by splitting these
distributions into the four different redshift bins
$0.2<z<0.4$, $0.4<z<0.6$, $0.6<z<0.8$ and $0.8<z<1.0$ for which we
have derived the HOD parameters from the CFHTLS. 

Since we are interested in the non-linear scales, in the following we
simulate the different surveys at the same sky coverage than the W1 field.
Doing
so, we can directly predict our
forecasted gain varying only the number of galaxies in the simulated
maps.
In summary, the redshift distribution enters our simulations in two
ways: first in the number of galaxies generated in the different maps,
secondly in the input angular two-point correlation via the Limber's equation \citep{Limber54}.

\begin{table*}
\centering
\caption{Specifications of the surveys that enter our prediction pipeline.}
\begin{tabular}{ccccccc}
\hline
\hline
Parameter& Description & DES & HSC & Euclid & LSST & WFIRST \\
\hline
$z_{m}$ & Mean redshift & 0.7 & 0.9 & 1.0 & 1.2 & 1.5 \\
$\bar{n}_{g}$ & Number density (arcmin$^{-2}$)& 10 & 22 & 30 & 50 & 95 \\
\hline
\hline
\end{tabular}
\label{table:surveys_carac}
\end{table*}

\section{Generation of mock galaxy and $A^{\ast}$ maps and spectra}
\label{sec:maps}
To produce simulations of a galaxy count map for a given set of cosmological and HOD parameters, we proceed similarly to
\cite{Carronetal14c}. The following summarize the different steps:
\begin{enumerate}
\item The galaxy two-point function $\omega(\theta)$ calculated according to \ref{subsec:HOD} is filtered to account for the cell finite size. We then obtain from relation~\ref{eq:xiAxid} the covariance matrix $\omega_A$ of the Gaussian field.
\item We generate the Gaussian field and exponentiate it to obtain the galaxy field. Note that the
  Fourier modes of the Gaussian map are not independent, since the finite volume breaks statistical translation invariance. Standard Fast Fourier Transform (FFT) based methods for Gaussian field generation are
  thus not applicable for our purposes. We use the ``circulant embedding''  method \citep[see][for details]{Carronetal14c}.
\item We then generate the count map $\vecN$ from the galaxy field and
  mask fractions, according to Equation~\eqref{sampling}, using a
  standard multinomial sampling algorithm. To take the
masks into account, we determine, in a Monte Carlo way, the effective
size for each cell which corresponds to the area that is contained in
the cell after subtraction of the masks. If the unmasked fraction $f$ of
the cell is less than a threshold of $0.3$, the cell is considered to
contain $\bar{N}$ objects. 
\item We estimate the angular averaged galaxy power spectrum by discrete Fourier transforming $\boldsymbol \delta_{g,i} = N_i/\bar N_i -1$, and averaging over the magnitude of Fourier modes
\beq
P_g(k) = \frac{1}{V}\frac{1}{N_k}\sum_{q \in \Delta(k)} \left |  \tilde{\delta_g} (\vecq)\right |^2,
\enq
where $N_k$ is the number of modes in the corresponding bin. In the
following, the notation $P(k)$ is used to designate the angular power spectrum
with $k=\ell + 1/2$. We use 20 $k$-bins equally spaced in $\ln k$, $k$ between 240 and 3100. The parameter $\bar N_{i}$ is defined as
\beq
\bar N_i  = f_i\lp \frac{  N_{\textrm{tot}}}{ \sum_{i} f_i}  \rp
\enq
We do not subtract any shot noise term, as this does not play a role
in the following (see appendix \ref{app:error}). 
\item Given a sampling rate $\bar N$, the mapping from $N$ to $A^*$ is defined by the non-linear equation \cite{Carronetal14a}
\beq \label{Astar}
A^* + \bar N \sigma^2_A e^{A^*}   = \sigma^2_A\lp N - \frac 12\rp,
\enq
where $\sigma^2_A = \ln (1 + \sigma^2_{\delta_g})$, with $\sigma^2_{\delta_g}$ the variance of the galaxy field fluctuations at the cell scale as predicted by the fiducial model. We transform the count map $\vecN$ to $\vecA^*$ solving that equation in each cell with sampling rate $\bar N_i$ with an efficient Newton-Raphson algorithm. The mean of the $\vecA^*$ map and its angular averaged spectrum $P_{A*}(k)$ is then extracted,
\beq
P_{A^*}(k) = \frac{1}{V}\frac{1}{N_k}\sum_{q \in \Delta(k)} \left |  \tilde{A^*} (\vecq)\right |^2.
\enq
\end{enumerate}

\begin{figure}
  \begin{center}
    \includegraphics[width=9cm]{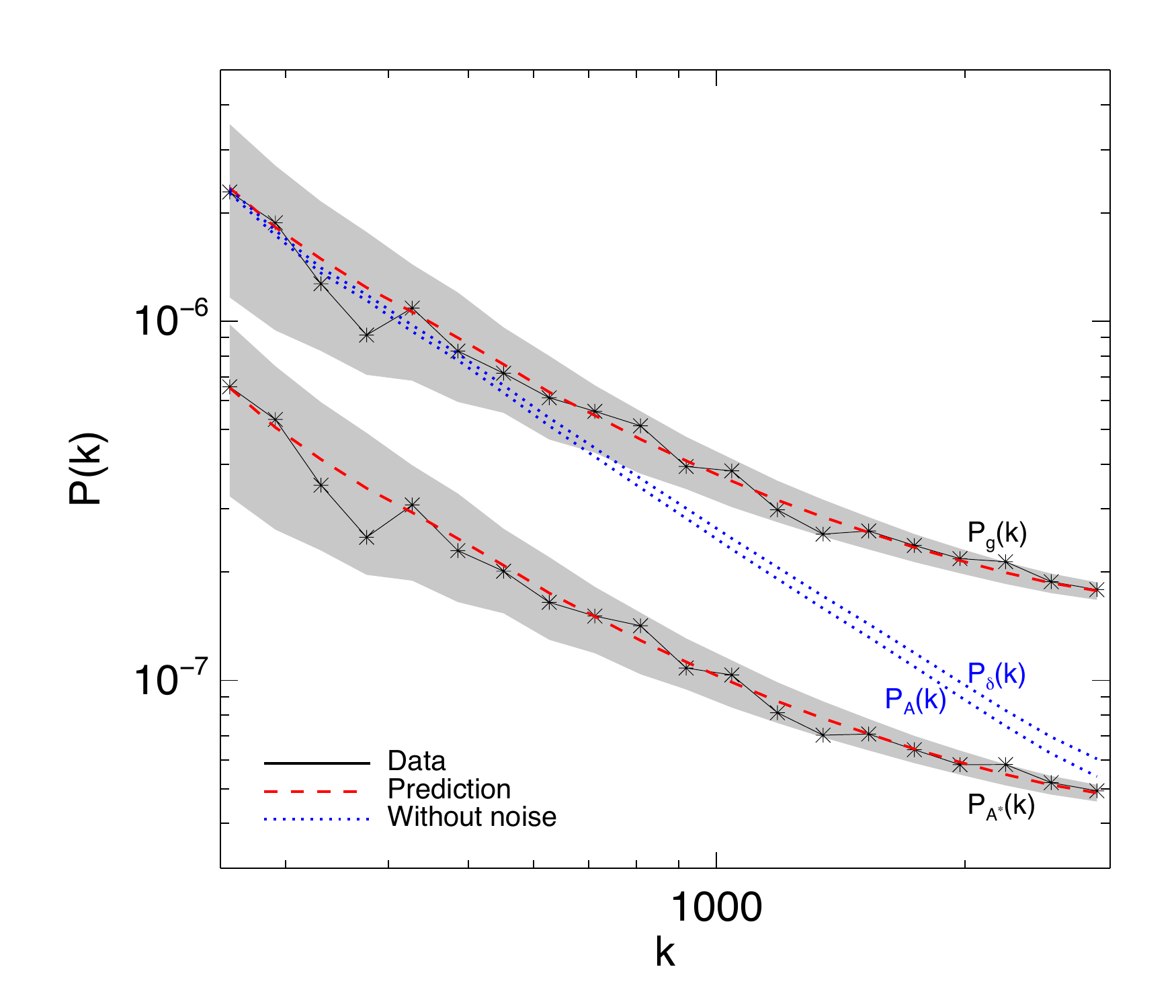}
\end{center}
\caption{Measurements of both the galaxy angular power spectrum
  $P_{g}(k)$ and the non-linear transform $A^{\ast}$ power spectrum $P_{A^{\ast}}(k)$
  on the CFHTLS W1 field in the redshift bin $0.6<z<0.8$ (solid black
  line). The red-dashed lines represent the predictions for both
  quantities derived as described in Section~\ref{sec:model} using
  $N_{res}=1000$ realizations. The grey area show the 2$\sigma$
  confidence regions. The blue dotted lines represent the power spectra of
  the underlying fields $\delta$ and $A=\ln(1+\delta_{g})$ and thus
  illustrate the effect of shot-noise on our predictions.}
\label{fig:Pkexample} 
\end{figure}

Figure~\ref{fig:Pkexample} shows the comparison between the measurement of the spectra in the CFHTLS data and predictions from the fiducial model obtained (together with their covariance matrices) with the help of a sufficient number of such simulations.
Shown are the spectra for the redshift bin $0.6<z<0.8$, as well as the 2$\sigma$
confidence regions (shaded).  The dashed line is the prediction from the fiducial model \citep[no fit was performed to these data points, the agreement only reflects that our simulation pipeline reproduces correctly the 2-point statistics of that data set as measured by][]{Wolk13}. A value of $\chi^{2} \sim 0.8$ in both
cases indicates that the model captures well the characteristics of
both the $A^{\ast}$-power spectrum and the covariance. In addition, the blue dotted lines 
correspond to the predictions of the power spectra for the
underlying fields $\delta_{g}$ and $A=\ln(1+\delta_{g})$. 

\section{Results}
\label{sec:est}
In this Section~\ref{sec:const}, we study the Fisher-matrix forecast gain of
using the $A^{\ast}$-power spectrum over of the galaxy power
spectrum. We use the following set of cosmological
parameters: the energy density of matter $\Omega_{m}$, the amplitude
of the power spectrum of initial conditions quantified in terms of
$\sigma_{8}$, and the dark energy equation of state parameter
$w_{0}$. Then in Section~\ref{sec:compdata}, we compare our
predictions to the actual gain measured from the CFHTLS data
in the same fashion.

\subsection{Fisher forecasts}
\label{sec:const}

Given a set of parameters \textbf{p}, the Fisher matrix provides a means to forecast the results of a likelihood analysis given
model predictions for a data vector \textbf{$\mathcal{O}$} and sample covariance
\textbf{C} of the observables.
The Fisher information matrix for Gaussian data with parameter independent covariance is given by:
\beq
F_{ij} = \frac{\partial \mathcal{O}^{T}}{\partial p_{i}}C^{-1} \frac{\partial
  \mathcal{O}}{\partial p_{j}}.
\label{eq:Fisherdef}
\enq
To a first approximation the inverse of the Fisher matrix
corresponds to the covariance of the posterior distribution of the
parameters one can obtain given the error bars on the data.
It means that the larger the value of a Fisher matrix coefficient is, the smaller the
variance becomes, and therefore, the tighter the constraints on the unknown parameter value.

In the case of the $N$-field the observable is the galaxy power spectrum
while for the $A^{\ast}$-field the observables are its mean and its spectrum.
Our goal in this paper is to quantify the gain using
the non-linear transform $A^{\ast}$
We study the information content of our observables by varying one parameter at the time among the set $(\Omega_{m}, \sigma_{8}, w_{0})$, comparing the values of the matrix \eqref{eq:Fisherdef}.
\newline
\indent
The covariance matrices for all surveys and all redshift
bins are estimated according to Section~\ref{sec:maps} using the
two-point correlation function at the fiducial parameters values and
$N_{res}=10,000$ realizations. The derivatives are calculated with
finite differences using $N_{res}= 5000$ realizations to estimate the ($A^{\ast}$-) power
spectra. Finally, we use
Equation~\ref{eq:Fisherdef} to calculate the Fisher matrix.
Table~\ref{tab:MatrixN} shows, as an example,
the Fisher matrices derived using on one hand the galaxy power spectrum
and on the other $A^{\ast}$ for the CFHTLS in the redshift bin
$0.6<z<0.8$. The use of the non-linear transform
$A^{\ast}$ provides more information on all the cosmological
parameters considered here. However, it is worth noting that $w_{0}$
is much less constrained than $\Omega_{m}$ and $\sigma_{8}$.

Considering Equation~\ref{eq:Fisherdef}, one can
easily see that the constraining power of a given data point depends
on the ratio between the derivative at this data point and its
errorbar: the larger the absolute value of the ratio is, the greater
the constraining power. Then, a crucial aspect in understanding our forecasts is to analyse how the different cosmological parameters
affect the galaxy clustering and then enter our predictions.

\begin{figure*}
  \begin{center}
    \begin{tabular}{c@{}c@{}}
      \includegraphics[width=5.5cm]{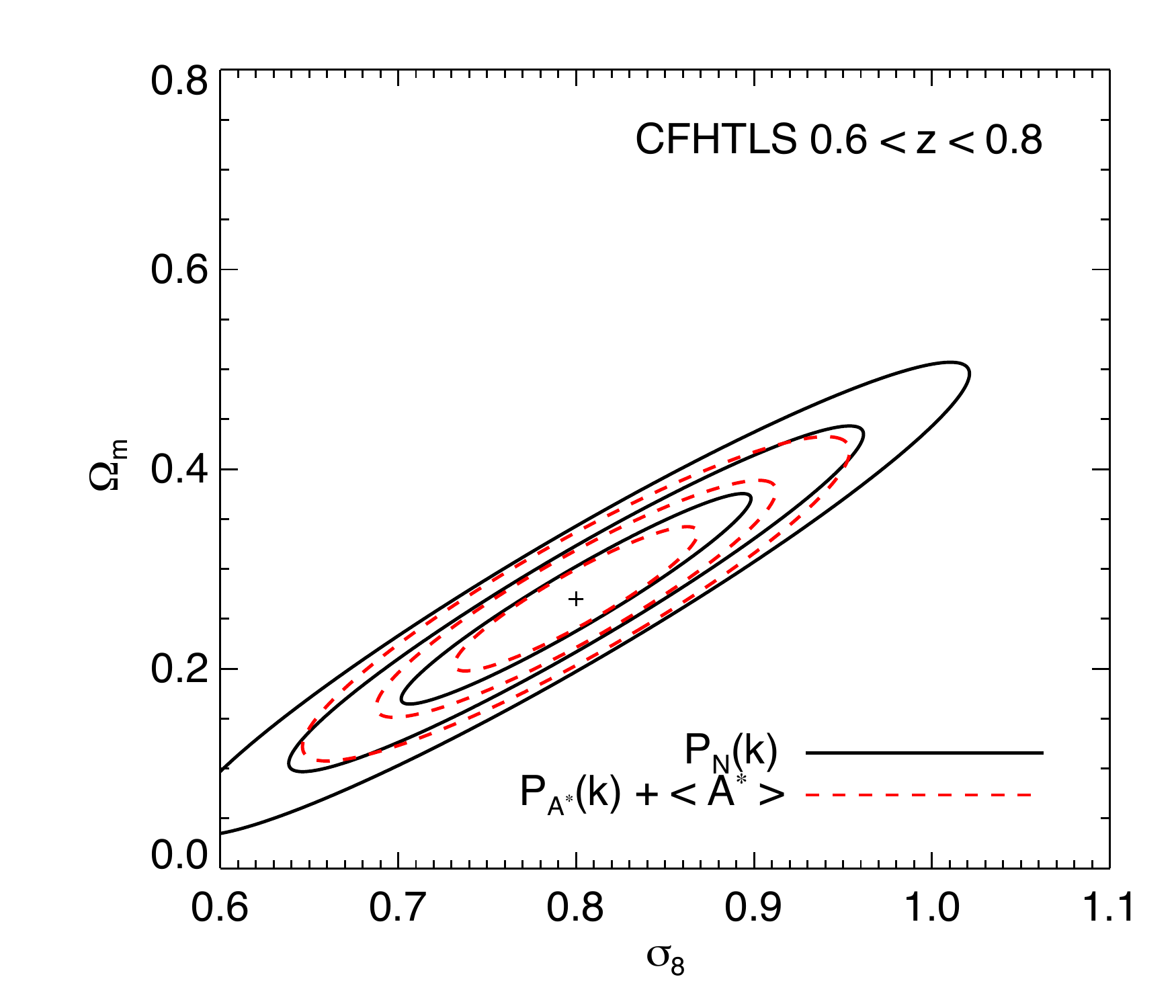}
      \includegraphics[width=5.5cm]{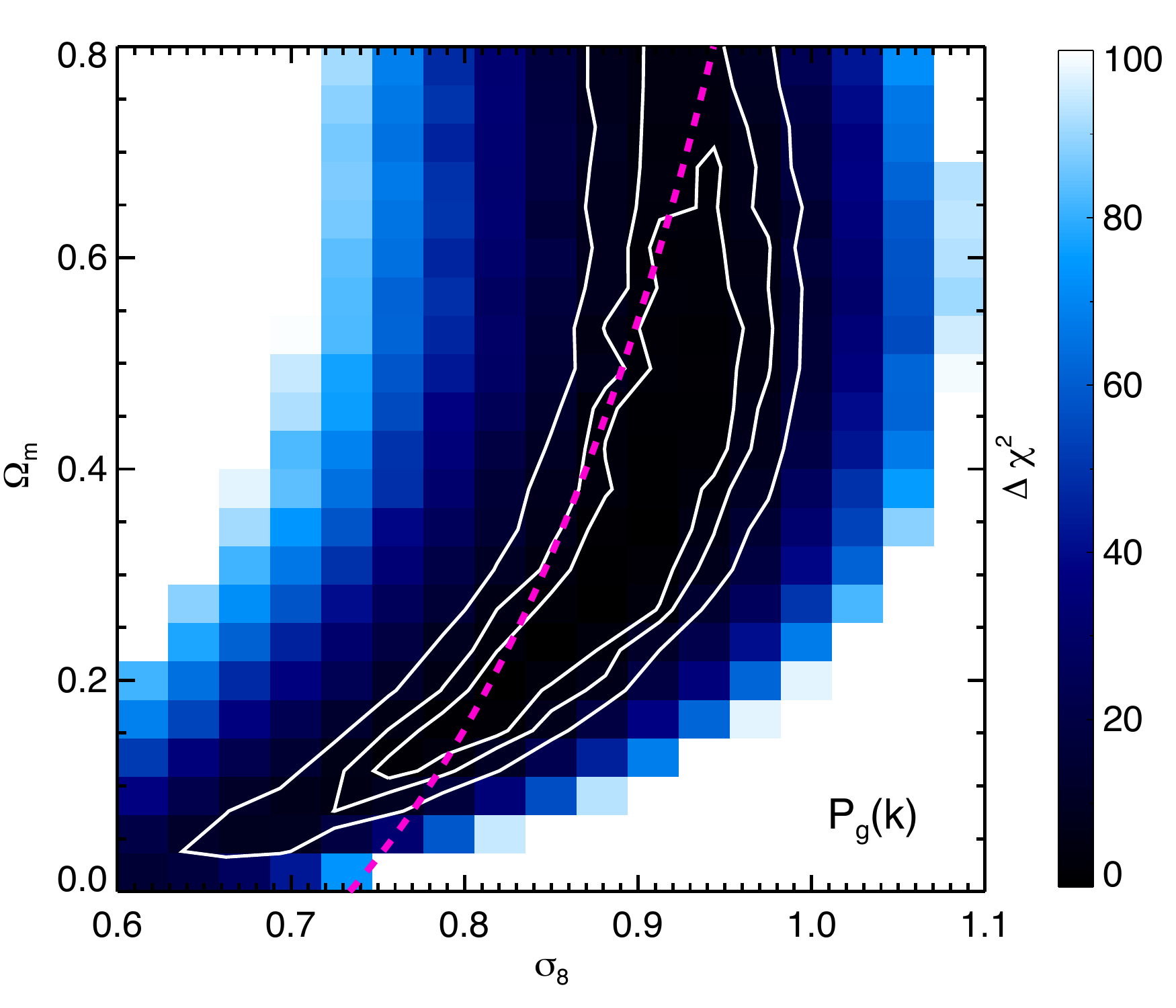}
      \includegraphics[width=5.5cm]{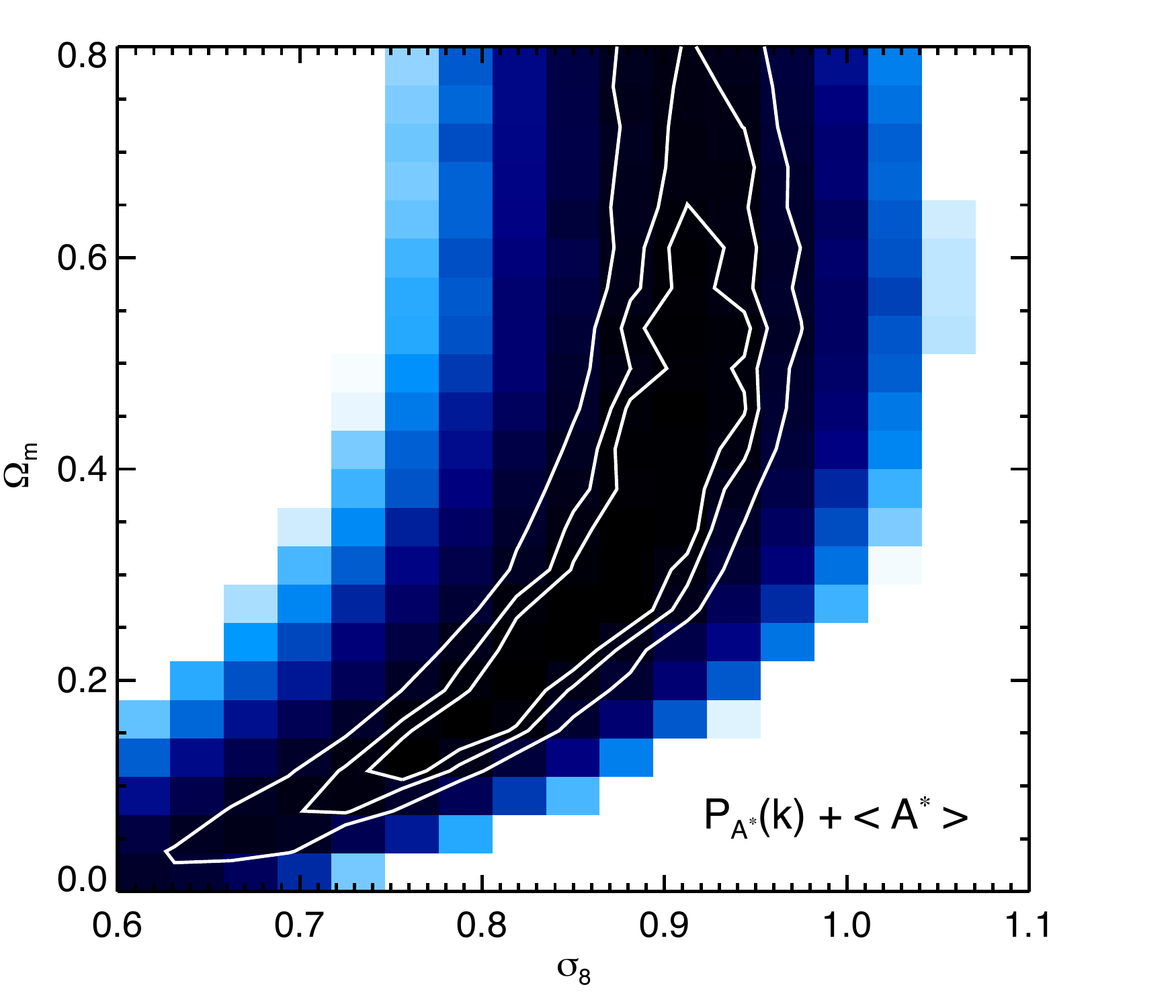}
      \end{tabular}
      \begin{tabular}{c@{}c@{}}
      \includegraphics[width=5.7cm]{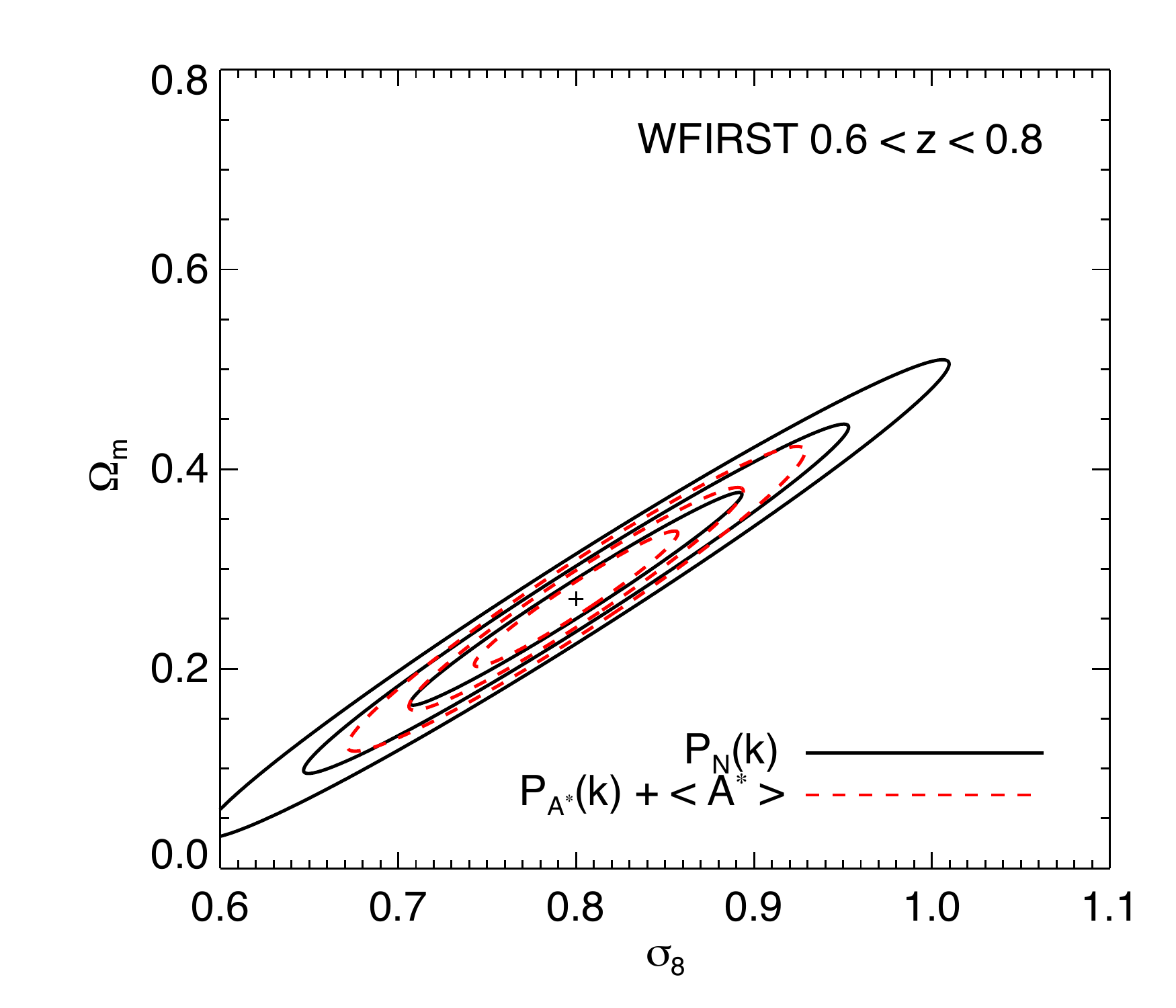}
      \includegraphics[width=5.5cm]{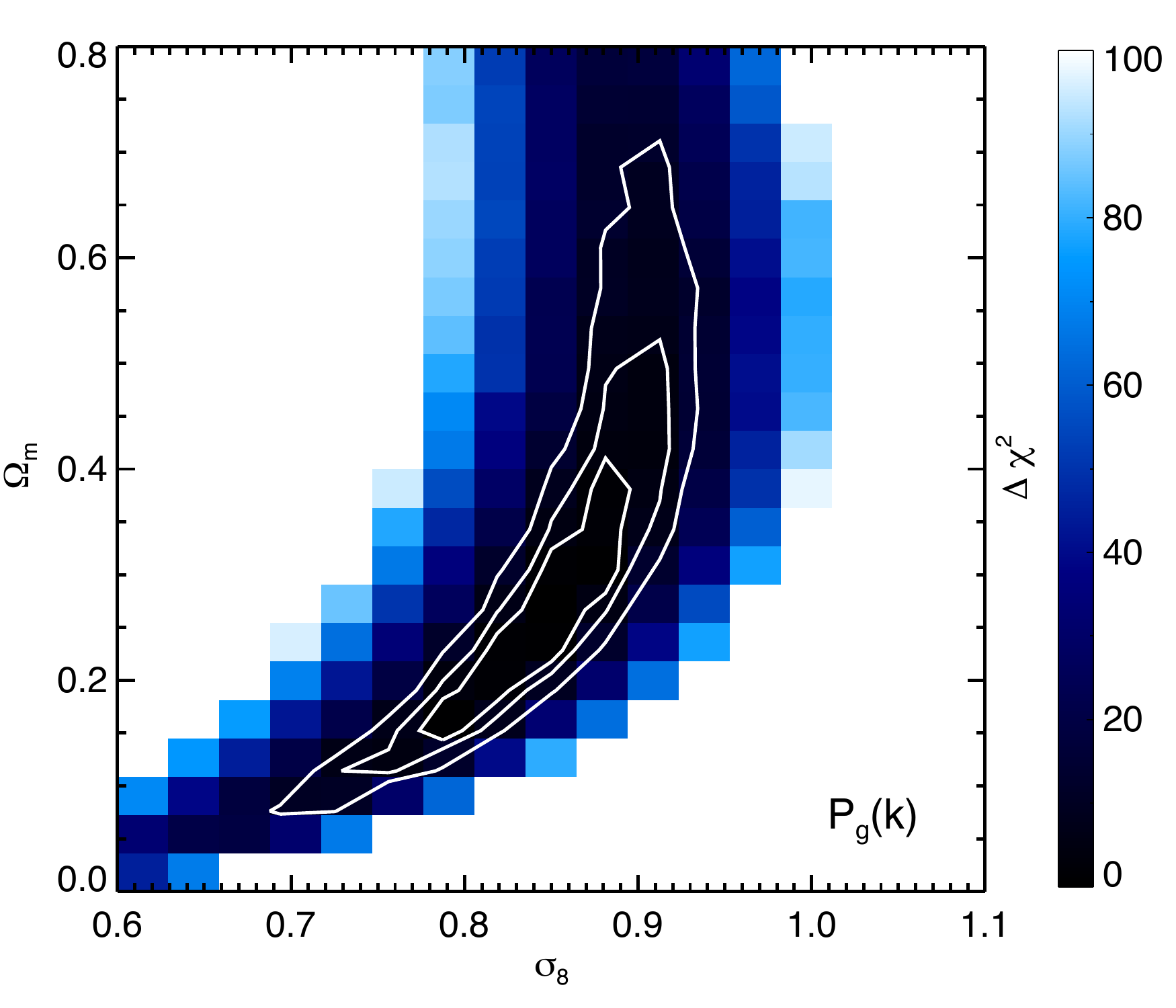}
      \includegraphics[width=5.5cm]{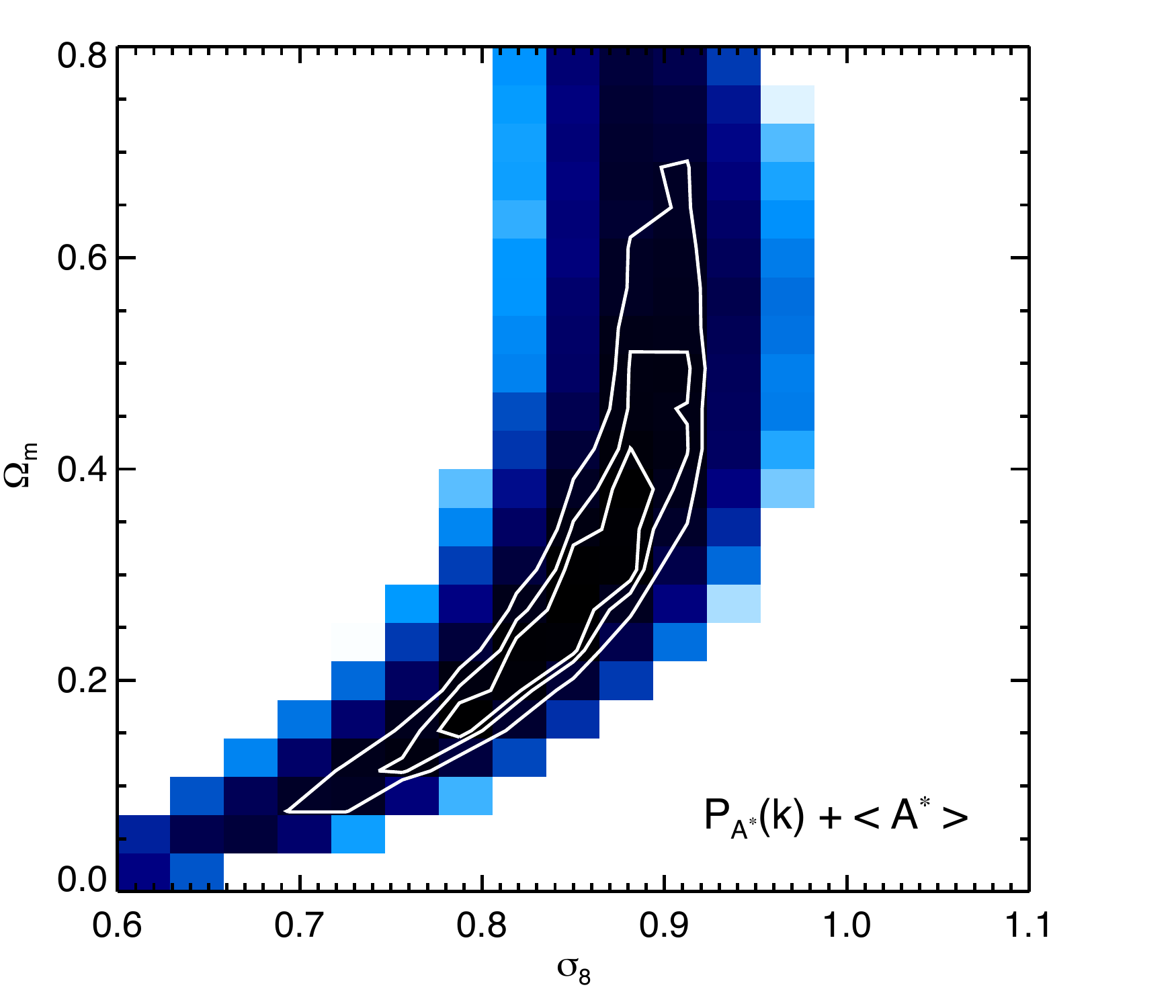}
      \end{tabular}
\end{center}
\caption{Comparison of the 68\%, 95\% and 99\%  confidence levels on
      $\Omega_{m}$ and $\sigma_{8}$. We take as illustrative example the bin
      $0.6<z<0.8$, both for the
      CFHTLS (upper panels) and the WFIRST (bottom panels). The upper and
      bottom right panels show the contours obtained using the power spectrum
      (solid black line) and $A^{\ast}$ (red
      dashed line) for both the CFHTLS and the WFIRST with a Fisher
      matrix analysis. The middle panels show the actual measurements for the
      confidence levels on the
      CFHTLS (top) and on a simulated map of the WFIRST (bottom) using
      the galaxy power spectrum. The dotted line illustrates the
      relation in Equation~\ref{eq:xipred}. The right panels shows the same
      measurements using instead $A^{\ast}$.}
\label{fig:forecastoms8}
\end{figure*}

The two-point correlation function of the galaxy field at large scales
can be written
as:
\beq
\omega  \propto [b \sigma_{8} D(a)]^{2},
\label{eq:xipred}
\enq
where $D$ is the growth factor and where the bias $b$ and $\sigma_{8}$
are kept fixed to their fiducial values.                 
Since increasing $\Omega_{m}$ implies more growth, keeping the value
of $\sigma_{8}$  fixed at $z=0$, leads to a decrease of the clustering strength at higher z.
For the same reason, increasing $\sigma_{8}$ results in an increase in the clustering strength.
Thus the derivatives with respect to
$\Omega_{m}$ and $\sigma_{8}$ have opposite signs, and the combination
$\Omega_{m}-\sigma_{8}$ is tightly constrained, while their sum is not. 
This can be seen on the upper left panel of the
Figure~\ref{fig:forecastoms8} which represents the comparison between
the forecast confidence levels on
$\Omega_{m}$ and $\sigma_{8}$ obtained using the galaxy power spectrum (solid
black line) and $A^{\ast}$ (red dashed
line). To quantify the improvement using the latter let us consider the 95\% confidence contours. The respective
areas for the two estimators are $0.027$ and $0.017$, corresponding to
a gain of about $1.26$ in the error bars.
As example of our gain expectations for an upcoming survey, the lower left panel of Figure~\ref{fig:forecastoms8} illustrates the
confidence levels for the WFIRST in the same redshift
bin. Unsurprisingly, the constraints on both $\Omega_{m}$ and
$\sigma_{8}$ are tighter compared to those from the CFHTLS.  The areas
of the 95 percent confidence contours are $0.0158$ and $0.008$ using
respectively the power spectrum and $A^{\ast}$, corresponding to a
gain of $1.37$ in the error bars.

\begin{figure*}
  \begin{center}
   \begin{tabular}{c@{}c@{}}
    \includegraphics[width=8cm]{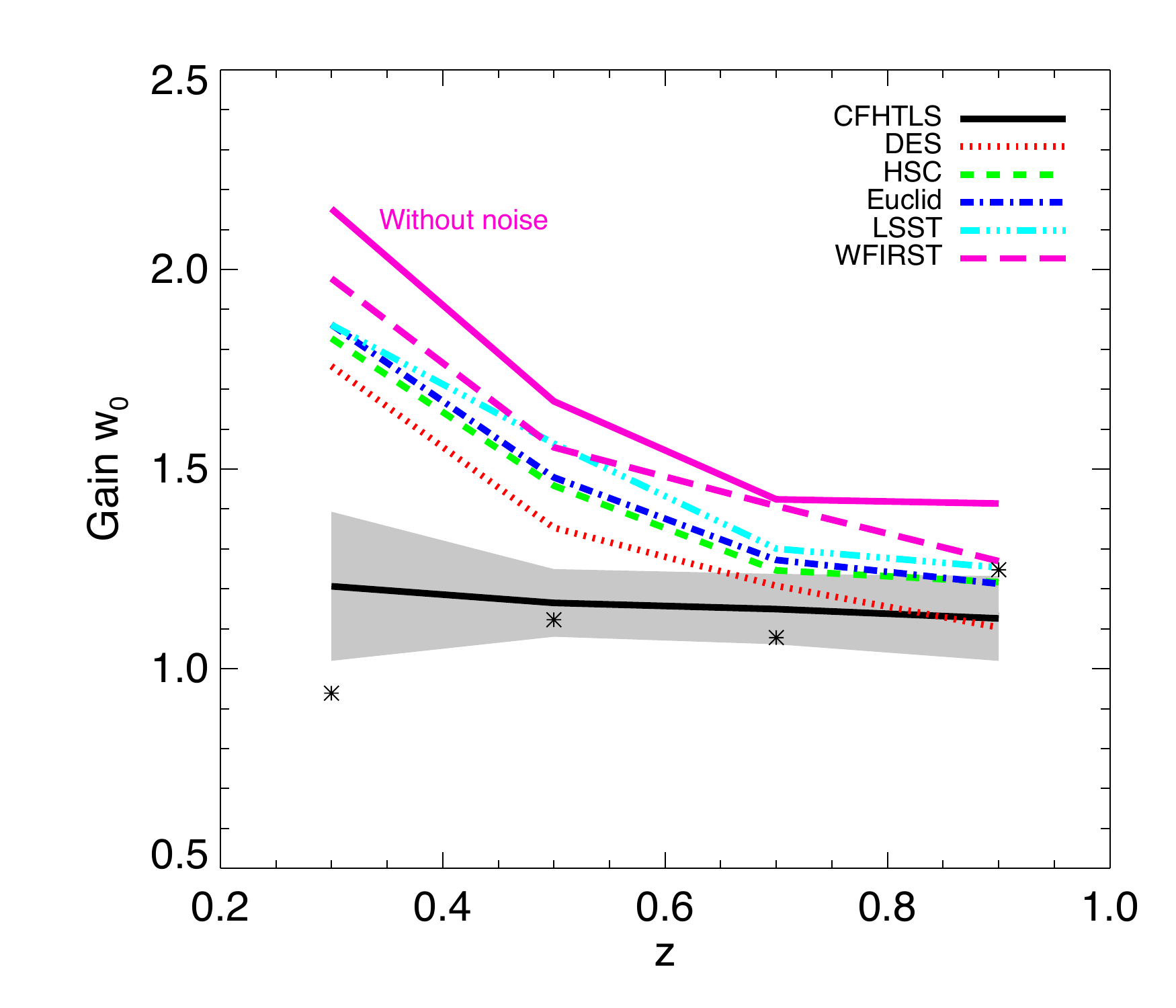}
    \includegraphics[width=8cm]{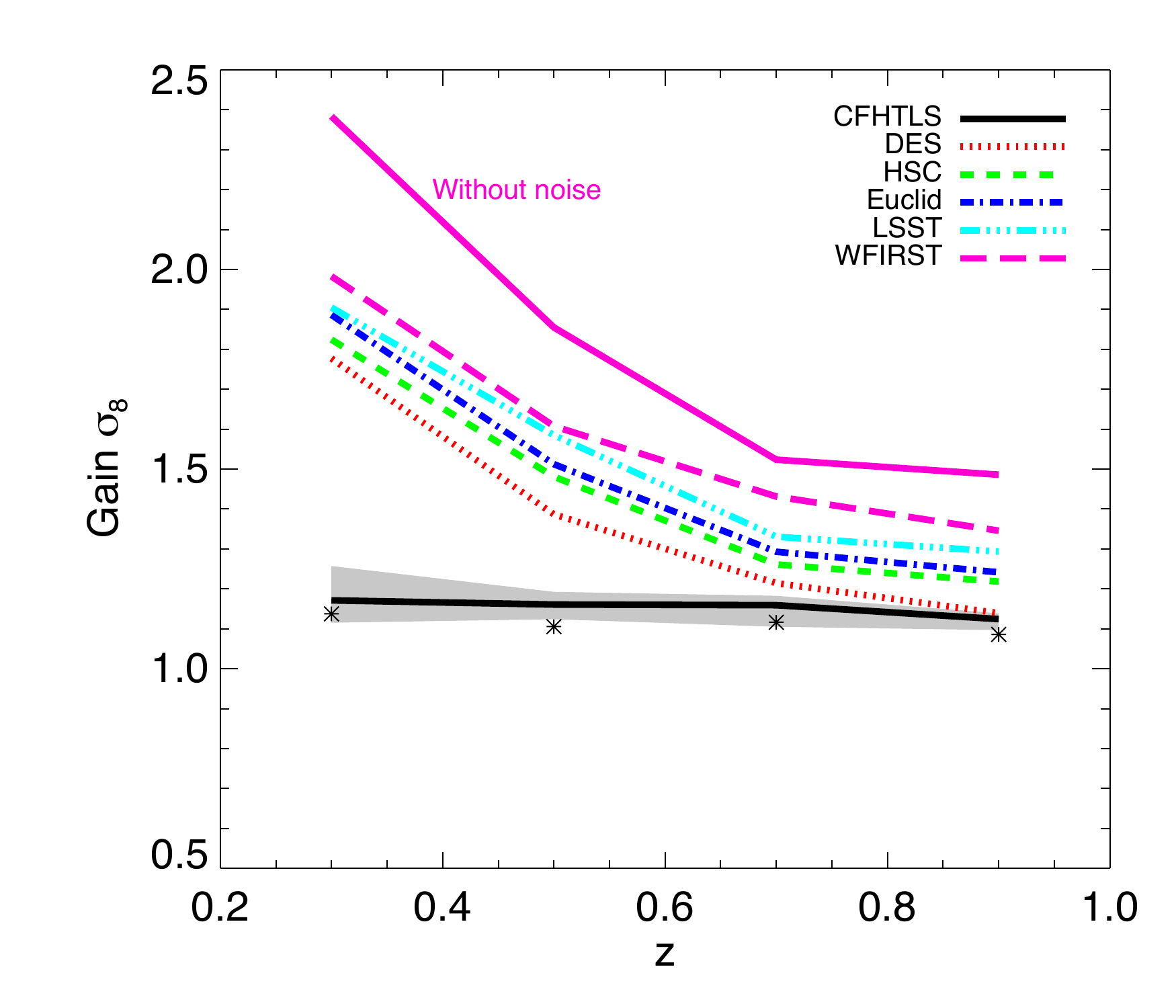}
  \end{tabular}
  \end{center}
  \caption{Forecast information gains on the
  cosmological parameters
  $w_{0}$ (left panel) and $\sigma_{8}$ (right panel) using $A^{\ast}$
  instead of the galaxy power spectrum for the CFHTLS (solid black line) and
  a list of future surveys in the redshift range $0.2<z<1.0$. The black stars
represent the comparison with actual measurements of this gain on the
CFHTLS data and the shaded areas show the 1$\sigma$ confidence limits
for the forecasts derived as described in
Section~\ref{sec:compdata}. The errors bars are larger in the case of
$w_{0}$ as this parameter is less constrained by the data. The upper
solid lines, in both panels, show the case with
no shot noise and thus represent the limit of the maximum information
that one can extract given our $\ell$-range.}
\label{fig:gain} 
\end{figure*}

Our main result is presented in Figure~\ref{fig:gain}, showing
the predicted improvement in the information on $w_{0}$ (left panel) and $\sigma_{8}$ (right
panel) for the upcoming surveys as a function of redshift. The quantity plotted is the ratio
between the $N$ and $A^{\ast}$ Fisher matrix elements, in that sense,
it represents the expected information gain using the
non-linear transform $A^{\ast}$ instead of the power spectrum. The simplest interpretation of this gain is an effective gain in survey area. The
gain for $\Omega_{m}$ follows closely the one for $\sigma_{8}$ as the
Fisher matrix coefficients are of the same order.

From Figure~\ref{fig:gain}, one can easily see that two different
trends emerge:
first using $A^{\ast}$ is more powerful at low redshifts where the
non-linearities are stronger, second
this new observable is more efficient for dense surveys such as the
WFIRST or the LSST. 
Table~\ref{tab:gain}
details for each survey the gain (maximum and average) that the new
method provides on the different cosmological parameters. In the particular case of the CFHTLS, the gain is only slightly above
1. Moreover, the error bars on the forecast gain for $w_{0}$ are
large due to the fact that this parameter is not well constrained and
thus diminishes the strength of the method. However, even for this less than
optimal case, considering the mean value of the gain,  $A^{\ast}$ performs better than the galaxy power spectrum over the whole redshift
range and thus unveils information otherwise hidden.
The upper solid lines in both panels show the predictions of the gain
without shot-noise. In that regime, $A^{\ast}$ reduces to the logarithmic
transform of the continuous field and therefore extract all the available
information on the cosmological parameters. These lines cannot be crossed and illustrate the upper
limit on the information given our chosen configuration.

All the upcoming surveys were designed to be  large, they are therefore  costly. 
The use of the optimal observable $A^{\ast}$ improves constraints to such a degree that
it corresponds to an effective increase of the survey area by up to a factor of 2.
Moreover, the largest gains are predicted at low redshift, exactly where
the dark energy is constrained the most efficiently. As a result, we
expect ``sufficient statistics'' to be a powerful method to
improve the future constraints on the various cosmological
parameters, just by using an alternative data analysis strategy.
    
\begin{table}
\centering
\caption{Fisher matrices for the CFHTLS-W1 field in the
  redshift bin $0.6<z<0.8$. The upper right coefficients and each of the first
coefficients on the diagonal are obtained using the galaxy
spectrum.  The lower left coefficients and the second
on the diagonal are obtained using $A^{\ast}$. The diagonal
elements are increased using $A^{\ast}$ while the off-diagonal elements
decrease.}
\begin{tabular}{cccc}
\hline
\hline
   & $\sigma_{8}$ &$\Omega_{m}$& $w_{0}$\\
\hline
$\sigma_{8}$ &  2517/2918 & -2234 & 868 \\
$\Omega_{m}$& -2518 & 2189/2614 & -794 \\
$w_{0}$              & 1005 & -911 & 303/351 \\
\hline
\end{tabular}
\label{tab:MatrixN}
\end{table}


\begin{table}
\centering
\caption{Information gain (max/average) for the different cosmological
parameters.}
\begin{tabular}{ccccc}
\hline
\hline
     & $\sigma_{8}$ & $\Omega_m$ & $w_{0}$ \\ 
\hline
 CFHTLS &        1.17/       1.15&        1.25/       1.17&        1.17/       1.16\\ 
 HSC &        1.82/       1.45&        1.70/       1.37&        1.83/       1.44\\ 
 Euclid &        1.89/       1.48&        1.78/       1.41&        1.86/       1.46\\ 
 DES &        1.78/       1.38&        1.64/       1.34&        1.76/       1.36\\ 
 LSST &        1.90/       1.53&        1.83/       1.44&        1.86/       1.49\\ 
 WFIRST &        1.98/       1.59&        1.87/       1.48&        1.98/       1.55\\
\hline
\hline
\end{tabular}
\label{tab:gain}
\end{table}

\subsection{Measurements from CFHTLS data}
\label{sec:compdata}
We can go further and compare our predictions to actual measurements
on the CFHTLS data.  We consider our fiducial model and use the W1 field in the four redshift bins $0.2<z<0.4$,
$0.4<z<0.6$, $0.6<z<0.8$ and $0.8<z<1.0$ to construct the galaxy counts
maps.

Figure~\ref{fig:forecastoms8}
shows the results obtained using a joint fit of the cosmological parameters $(\Omega_{m},
\sigma_{8})$ for both the galaxy power spectrum (middle panel)
  and the $A^{\ast}$-power spectrum (right panel). The measurements
  are made in the redshift bin $0.6<z<0.8$ and our predictions are
  derived with $N_{res}=100$. One can recognize for the two observables the usual ``banana'' shape for the
contours on $\Omega_{m}$  and $\sigma_{8}$. This shape is roughly
reproduced by the dotted line which is given by Equation~\ref{eq:xipred}. This shows that the large scale behavior of the two-point
correlation function can explain the general trend of our
measurements but also that small scale contributions enter our predictions
and result in more subtle effects.
The contours represent,
as before, the 68\%, 95\% and 99\% confidence levels.
As a comparison, we perform the same analysis on a simulated map for
the WFIRST survey also in the redshift bin $0.6<z<0.8$. This is shown on the
bottom middle and right panels of Figure~\ref{fig:forecastoms8}. We
see the ``banana'' shape for the contours of the galaxy power spectrum
and $A^{\ast}$, we also observe that using the latter the contours on
$\Omega_{m}$  and $\sigma_{8}$ shrink, leading to better constraints
on these two parameters in a way that mirrors well the
predictions made before using the Fisher forecast.

\begin{figure*}
  \begin{center}
    \begin{tabular}{c@{}c@{}}
      \includegraphics[width=5.5cm]{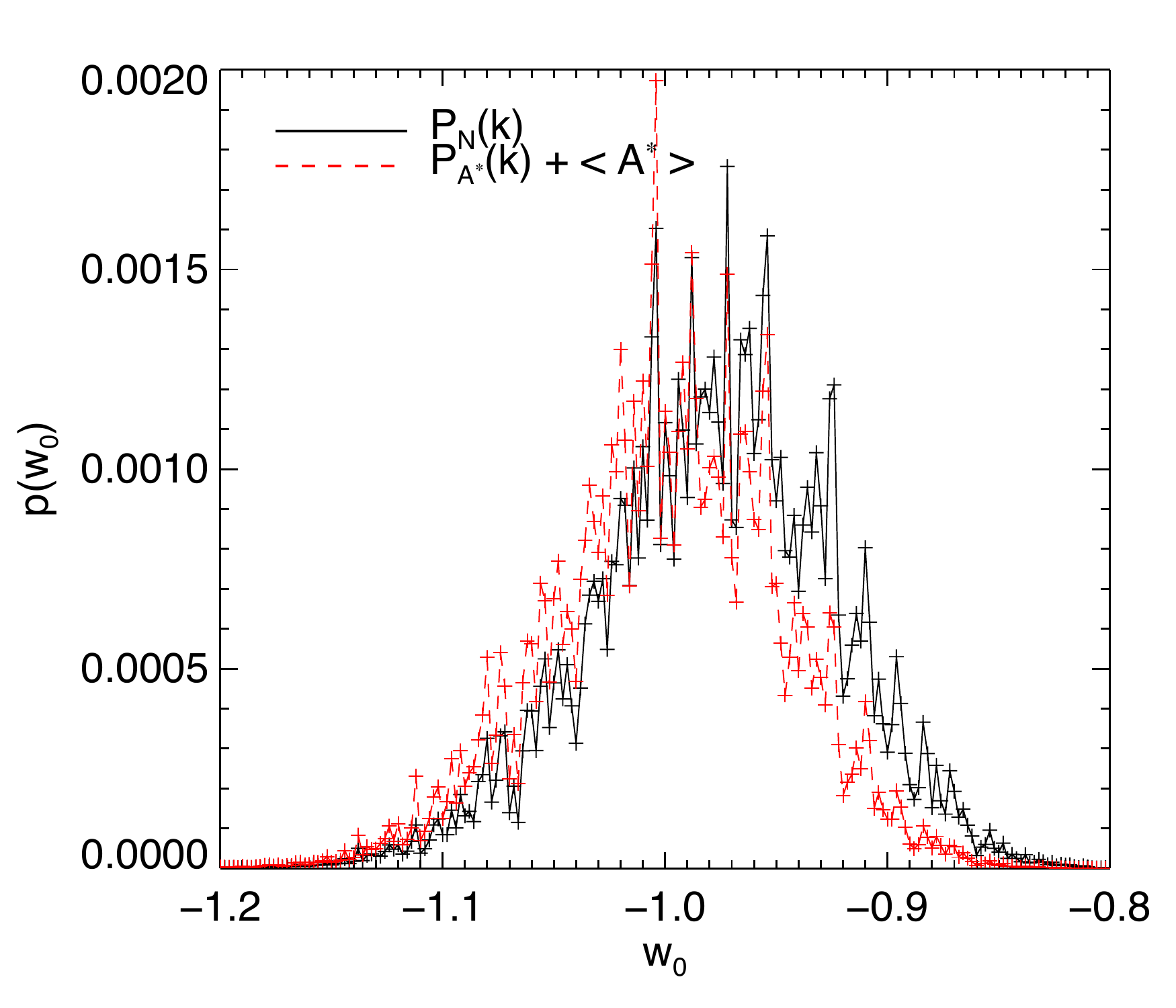}
      \includegraphics[width=5.5cm]{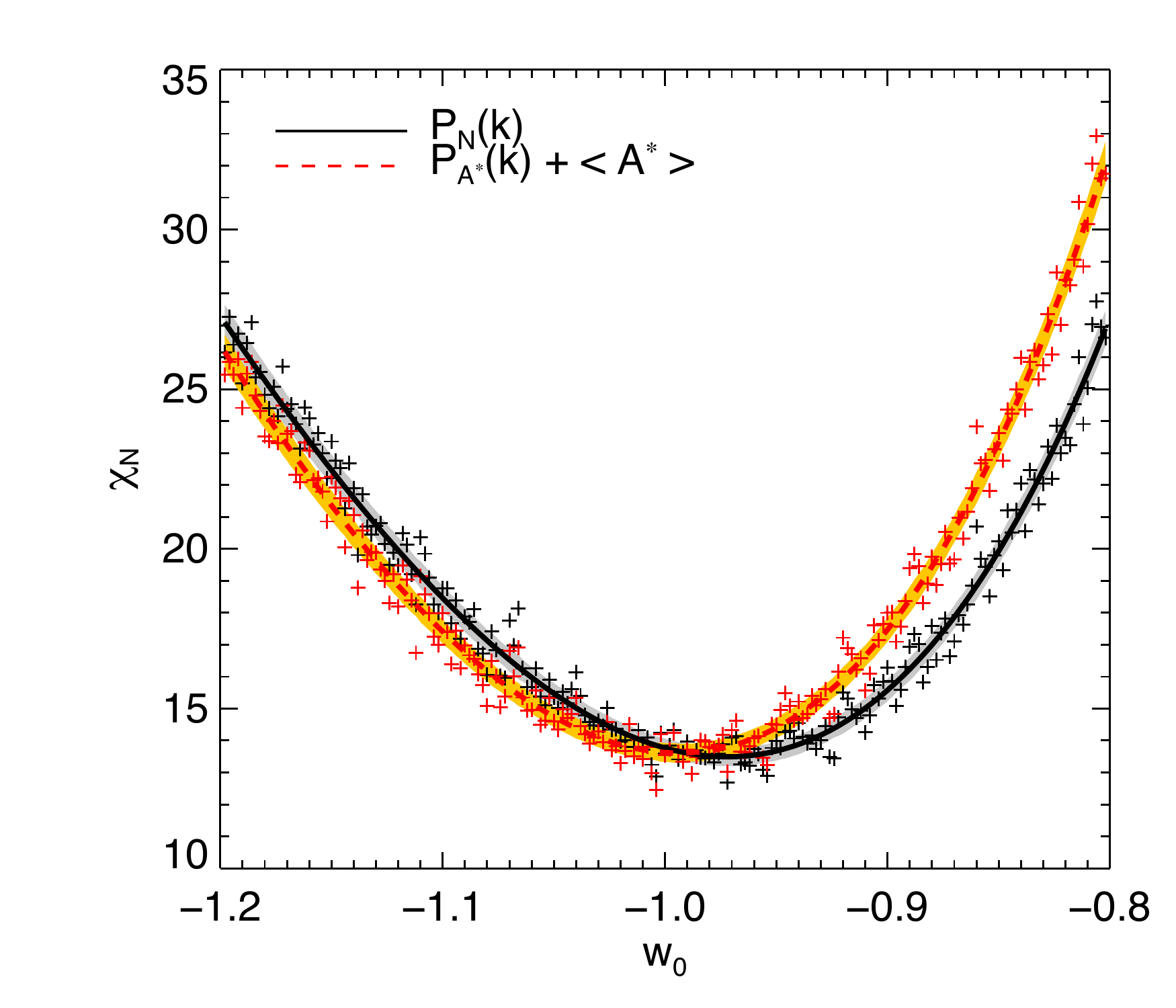}
      \includegraphics[width=5.5cm]{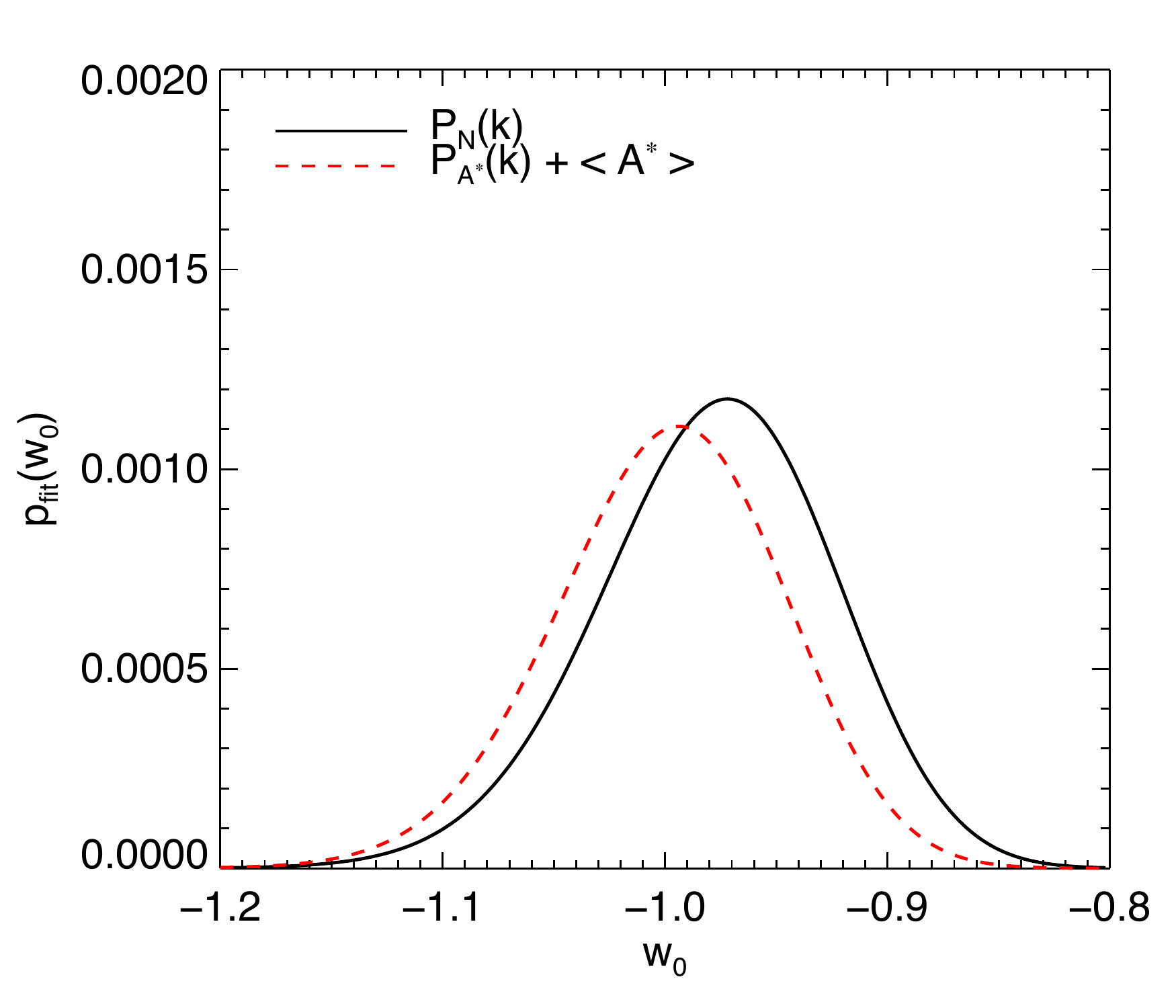}
  \end{tabular}
\end{center}
\caption{An example to illustrate the procedure to estimate the variance of
  the posteriors. Here we fit the
  cosmological parameter $w_{0}$ on the CFHTLS data in the redshift
  bin $0.6<z<0.8$. The left panel show the noisy posteriors obtained
  using the galaxy power spectrum (solid black line) and $A^{\ast}$ (red dashed line). To have an estimate of the
variances of these posteriors, we fit $\chi^{2}_{N}$ with a 4-th order
polynomial using the error bars described in Appendix~\ref{app:error}. The middle panel illustrates this procedure and show the best
fit overplotted to the measurements (red and black symbols). The right
panel shows the ``smoothed'' posterior estimated from the best fit
values for both the galaxy power spectrum (solid black line) and $A^{\ast}$ (red dashed line).
}
\label{fig:methoderr}
\end{figure*}

We can also compare the expected gains on Figure~\ref{fig:gain} for the
two parameters $\sigma_{8}$ and $w_{0}$ to the gain directly measured on
the data.
We proceed as followed:
\begin{enumerate}
\item For each redshift bin we construct the galaxy counts
maps of W1.
\newline
\item Fixing everything else and varying one cosmological parameter at
  a time, we fit our predictions for both the galaxy power spectrum
  and the $A^{\ast}$-power spectrum to
  the ones measured on the maps. We use a $\chi^{2}$-technique on a $200$-points grid going
  from $[-2.0, 0]$ for $w_{0}$ and $[0.4, 1]$ for $\sigma_{8}$
  respectively with $N_{res}=200$.
\newline
\item Our simulations are made with a finite number of realizations
  which results in noisy posteriors (as it can be seen on the first
  panel of
  Figure~\ref{fig:methoderr}). It is thus not clear how to estimate
  their variances directly. However, the estimated ``chi-squared'',
  $\chi^{2}_{N_{res}}$, is a well-behaved function as its errors
  bars are symmetric around the mean value as shown in
  Appendix~\ref{app:tech}.  We fit to its values a 4-th order polynomial which gives us
  an analytical expression for $\chi^{2}_{N_{res}}$ and therefore for a
  ``smoothed'' version of the posteriors. If the
  posteriors were Gaussian, they will be described by a 2-th order
  polynomial. These steps are summarized in Figure~\ref{fig:methoderr}.
\newline
\item We then estimate the variances of the ``smoothed''
  posteriors obtained using
  both $A^{\ast}$ and the galaxy power spectrum and finally take the ratio of the
  two to measure the gain.
\end{enumerate}

The results are represented by the black stars on
Figure~\ref{fig:gain}. We see some discrepancies compared to the
forecast value of the gain especially for the less constrained
parameter $w_{0}$. To discriminate between a tension or just an effect
of the variability in our data, we estimate the uncertainties on the
forecast gain by repeating the steps above for 100 simulated data maps of W1.
We then use the variance of the obtained ratios as an
  estimator of the uncertainties on our forecast gains which are
 illustrated by the 1$\sigma$ confidence regions (the shaded areas) in Figure~\ref{fig:gain}.
We see that within the error bars the overall agreement
with the predictions is good meaning that our statistical model
captures and reproduces accurately the behavior of the data.
All of the above demonstrates that the prediction pipeline implemented
in that study describes precisely the statistical properties of the
data and therefore shows that our predictions for the upcoming surveys
and the expected gain for $A^{\ast}$ are both realistic and robust.

\section{Discussion and prospects}
\label{sec:discuss}
It was known that non-linear transforms help to capture more
efficiently the information encoded in the matter density field. In
this work, we have shown, in a quantitative way, that there
is room from improvement beyond the galaxy power spectrum for the
clustering of the large scale structures using a new observable
$A^{\ast}$ derived to be the ``sufficient statistics'' in the case of
the galaxy field.
We have developed a simulations pipeline which include all the main
sources of statistical uncertainties (super survey modes, galaxy trispectrum,
discreteness effects) and calibrated our modelling on actual data
coming from one of the state-of-art large photometric redshift
surveys available at the time: the CFHTLS. From this pipeline we were able
to simulate, using a large number of realizations, both the galaxy and
the $A^{\ast}$ power spectra.

We have demonstrated that our statistical modeling is
accurate and captures correctly the statistical properties of the
measurements. We have compared the efficiency forecast for the galaxy power
spectrum and the mean and spectrum of $A^{\ast}$ for the CFHTLS to
measurements on this data set. In this particular case, the gain using the non-linear
transform is modest, especially for parameters that are not well
constrained by the data. The promise of the new observable is larger for
upcoming surveys. We found that with higher signal to noise, the
gain on the information on the three cosmological parameters $\Omega_{m}$,
$\sigma_{8}$ and $w_{0}$ is up to about a factor of 2, especially at low
redshifts and for dense surveys.

During the  practical implementation of this
estimator, we needed to solve some
technical difficulties that are unique to this method.
First, as we used a finite number of realizations to predict the power
spectrum, the obtained
posteriors for the different parameters are noisy. We
described in Appendix~\ref{app:tech} that our errors on the parameters
  even if noisy converge very fast needed only a reasonable amount of
  realizations and we suggested a practical method to obtain smooth
  posteriors. Second, for the same reasons, this method, as it mirrors
  the statistical uncertainties of the measurements, works better for
  parameters that are well constrained; this is a perfect fit for high precision
  cosmological applications. 
  

Despite that the technology of estimation is admittedly slightly more complex (and we provided a detailed description of complexities and how to mitigate them), we have shown that ``sufficient
statistics'' increase the statistical constraining power of upcoming surveys to the point that the additional effort is worth the consideration.
 In a future work, we will use this new optimal observable to put simultaneous
constraints on both the HOD and the cosmological parameters, testing
the effects of priors and of the combination of different independent
measurements, e.g., in combination with Planck \citep{Planck14}.
Another natural extension of $A^{\ast}$ will be to derive the
``sufficient statistics'' for the galaxy shear field.  While this is
less straightforward due to the
mass-sheet degeneracy \citep[][e.g.]{Carronetal14b}, in combination 
with additional measurements, such as the galaxy clustering and the
galaxy-galaxy lensing, it will provide optimal constraints on cosmological parameters.

\section{Acknowledgments}
The authors
thank Suhrud More, Masahiro Takada and Alexie Leauthaud for useful
conversations.
\newline
\indent
The authors acknowledge NASA grants NNX12AF83G and NNX10AD53G for support.
\newline
\indent
Part of this work was based on observations obtained with MegaPrime/MegaCam, a joint project of CFHT and CEA/IRFU, at the Canada-France-Hawaii Telescope (CFHT) which is operated by the
National Research Council (NRC) of Canada, the Institut National des
Science de l'Univers of the Centre National de la Recherche
Scientifique (CNRS) of France, and the University of Hawaii. This work
is based in part on data products produced at Terapix available at the
Canadian Astronomy Data Centre as part of the Canada-France-Hawaii
Telescope Legacy Survey, a collaborative project of NRC and CNRS.

\begin{appendix}
\section{Estimation of the likelihood with a finite number of
  simulations}
\label{app:tech}
We use a finite number of simulations to obtain the model predictions at each point in parameter space. 
We discuss in this section how this impacts the calculation of the parameter posteriors, allowing the assessment of the 'errors on the parameters errors'. We find that this effect broadens only slightly  the width of the parameter posterior according to Equation~\eqref{broadenchi2}.
\newline
\newline
The likelihood for the model parameters is proportional to $e^{-\chi^2/2}$, where
\beq
\chi^2(\vectheta) = \lp \vecd - \bar \vecd(\vectheta) \rp\cdot \Sigma^{-1} \lp \vecd - \bar \vecd(\vectheta) \rp,
\enq
where $\vecd$ is the data vector (in our case the spectra and/or mean of $\vecA^*$) and $\bar \vecd$ the predictions. In this paper, we evaluate the predictions at each point in parameter space by averaging over $N_{res}$ simulations of the data. Therefore, the true $\chi^2$ is estimated with some error and bias by
\beq
\hat \chi^2_{N_{res}} (\vectheta) =  \lp \vecd - \hat \vecd_{N_{res}}(\vectheta) \rp\cdot \Sigma^{-1} \lp \vecd - \hat \vecd_{N_{res}}(\vectheta) \rp,
\enq
where $\hat \vecd_{N_{res}}$ is the average over $N_{res}$ simulations of the data vector. Since the simulations are independent, the central limit theorem implies that for reasonably large $N_{res}$ the estimate $\hat d_{N_{res}}$ will be a Gaussian vector, even if $\vecd$ is not. The mean of $\hat \vecd_{N_{res}}$ is $\bar \vecd$ and its covariance matrix $\Sigma / N_{res}$. The PDF for $\hat \chi^2_{N_{res}}$ can be given in closed form (it is basically a non-central $\chi^2$ variable). Again, it will be for all practical purposes a Gaussian. Its bias with respect to the true $\chi^2$ and its variance can be straightforwardly calculated from the above expression, with the result
\beq \label{chi2var}
\av{\hat \chi^2_{N_{res}}} - \chi^2 = \frac{N_d}{N_{res}},\quad \Var{\hat \chi^2_{N_{res}}} = \frac{2N_d^2}{N_{res}^2} + \frac{4}{N_{res}} \chi^2,
\enq
where $N_d$ is the dimension of the data vector.
Thus, the estimated (unnormalized) parameter likelihood 
\beq \label{pn}
\hat p_{N_{res}} \propto \exp{\lp - \hat \chi_{N_{res}}^2(\vectheta)/2 \rp}
\enq is the exponential of a Gaussian variable, i.e. a lognormal variable at each point in parameter space.
Furthermore, the estimates at different points are independent. With this at hand, we can then ask how well we can measure some properties of the parameter posterior. Assuming the prior does not play a role, estimates of some function $f(\vectheta)$ such as the mean or variance of the posterior read on average
\beq
\av{ \frac{\int d\vectheta f(\vectheta) \hat p_{N_{res}}(\vectheta)}{\int d\vectheta  \hat p_{N_{res}}(\vectheta)} } = \frac{\int d\vectheta f(\vectheta) \av{ \hat p_{N_{res}}(\vectheta)}}{\int d\vectheta \av{ \hat p_{N_{res}}(\vectheta)}}. 
\enq
Fluctuations from that relation decaying away with the number of points with which the likelihood is sampled. With Equation~\eqref{pn}, Equation~\eqref{chi2var} and the fact that $\av{e^x} = e^{\av{x} + \Var x /2}$ for Gaussian $x$, the expectation value of $\hat p_{N_{res}}$ can be calculated with uncomplicated algebra. The result is very simply
\beq
\av{\hat p_{N_{res}}(\vectheta) }\propto \exp \lp -\frac 12 \chi^2(\vectheta)\lp 1 - \frac 1 N_{res}\rp\rp.
\enq
The correction is due to the second term in the variance in Equation~\eqref{chi2var}, the first term and the bias in that equation being absorbed in the normalization constant. We can conclude that the posterior is only slightly homogeneously broadened. In particular, if the true posterior is roughly Gaussian, the inference parameter (co)variance is slightly larger,
\beq \label{broadenchi2}
\frac{\sigma_{\vectheta}^{\textrm{inferred}}}{\sigma_{\vectheta}^{\textrm{true}}} = \sqrt{\frac{N_{res}}{N_{res}-1}}.
\enq
\begin{figure*}
\begin{center}
\includegraphics[width = 0.45\textwidth]{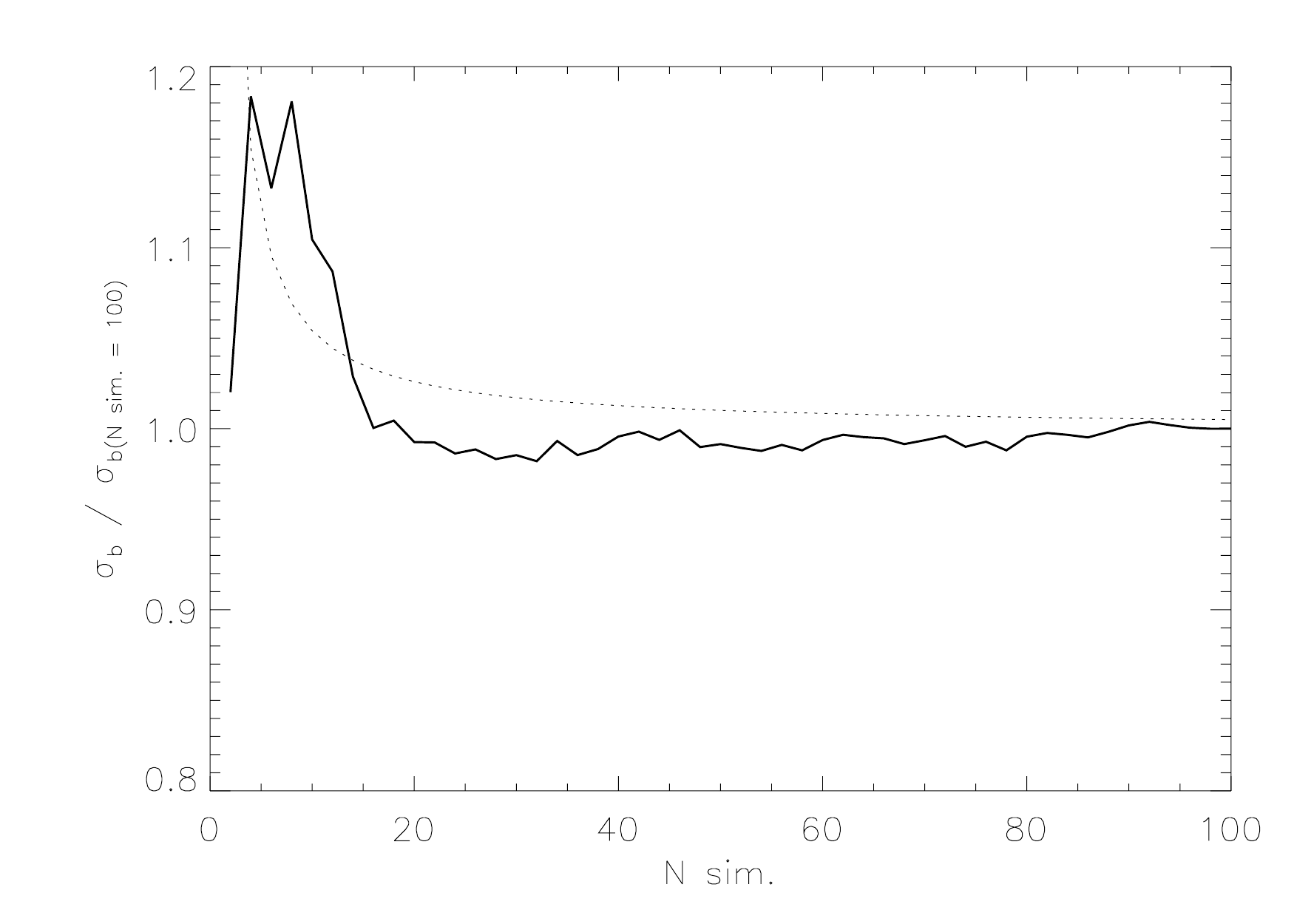}
\includegraphics[width = 0.45\textwidth]{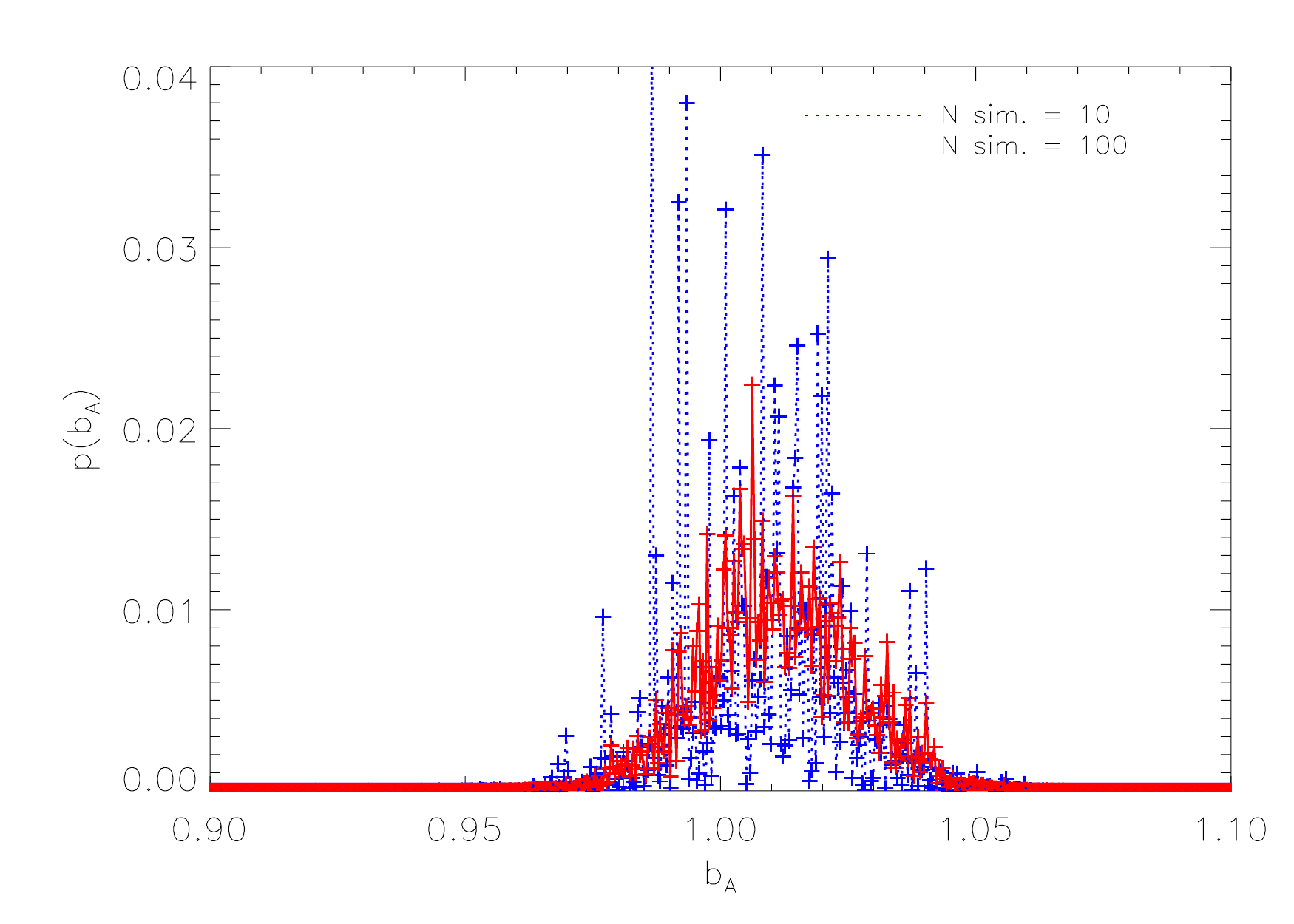}
\caption{A test case showing the evaluation of the parameter posterior (here for a bias parameter in the log-density) from the extraction of the spectrum $P_g(k)$, as a function of the number of simulations at each point in parameter space used to obtain the model predictions for the spectrum. The lower panel shows the posterior for $N = 10$ and 100 simulations. The upper panel shows as a function of $N$ the square root of the variance of the posterior, normalized to its value found for $N = 100$. Since the wild fluctuations in the posterior are uncorrelated, the variance (or other averages) takes a well-defined value even for moderate $N$. The dashed line on the upper panel is $\sqrt{N/(N-1)}$, Equation~\eqref{broadenchi2}, predicted by the simple arguments in this section.}
\label{figtest}
\end{center}
\end{figure*}
These results are illustrated in Figure \ref{figtest}.

\section{Multinomial versus Poisson sampling}
\label{app:error}
\begin{figure}
  \begin{center}
    \includegraphics[width=9cm]{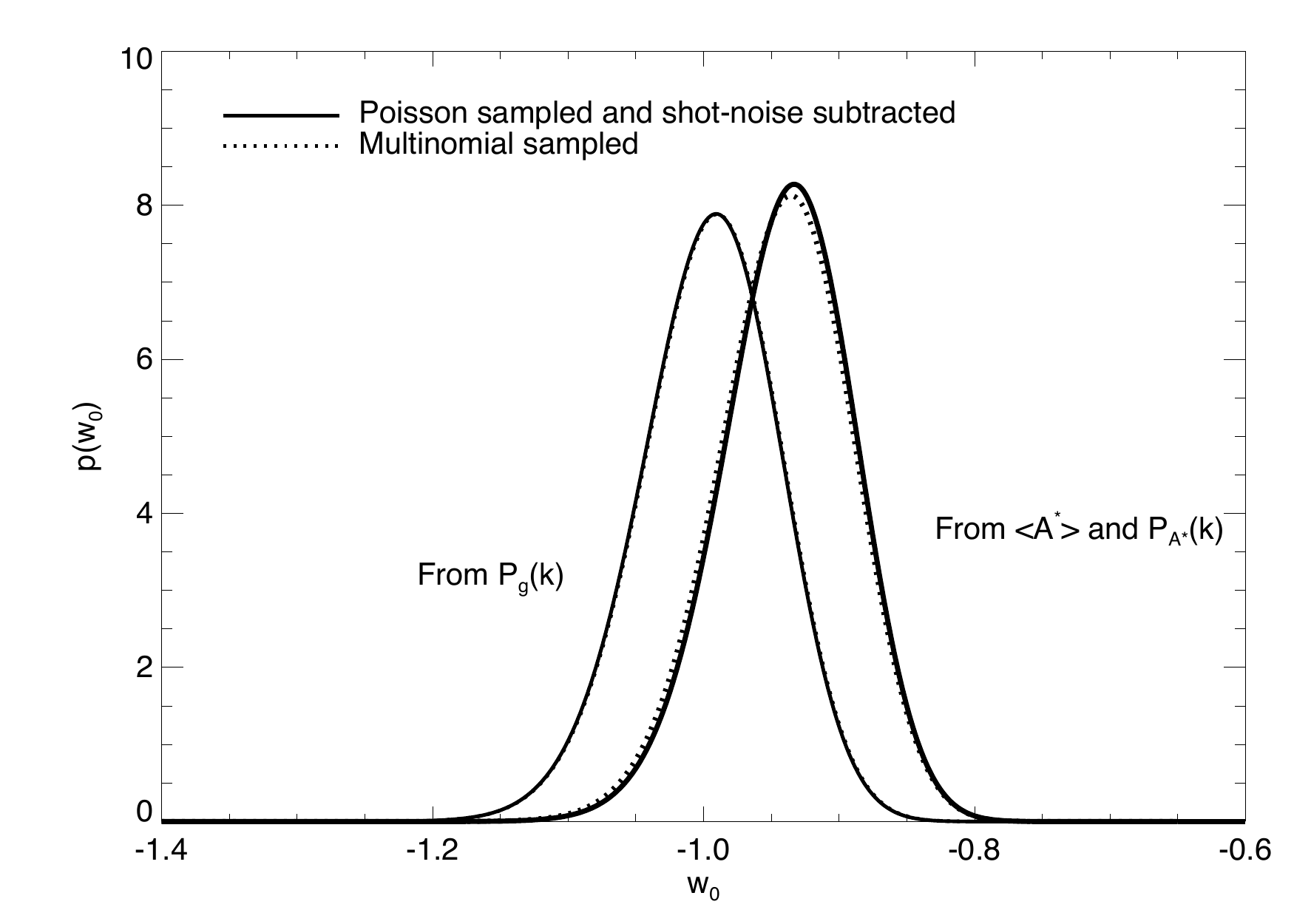}
  \caption{\label{figure:PoissonVSmulti}  An illustration of the equivalence of multinomial (used in the paper) and Poisson sampling for our purposes. The figure shows the posterior of the dark energy equation of state $w_0$ obtained by fitting the spectrum of $\delta_g$ or the mean and spectrum of $A^*$. The dotted line shows the constraints using Poisson sampling, which requires the introduction of a sampling intensity parameter, its estimation from each simulated map and subtraction of the estimated shot noise component. Multinomial sampling shown as the solid lines lead to the same results with no need for any of these steps.}
\end{center}
\end{figure}

Two natural choices of discrete sampling of the underlying field to
represent the projected counts are Poisson sampling, where the number
of galaxies in each cell field is drawn from a Poisson distribution,
and multinomial statistics, where a fixed number of galaxies in
distributed throughout the map.  In this paper we chose the second
option, mainly for its simplicity. We describe here some of the
differences between the two choices, and show that choosing
multinomial or Poisson will not change our results.
\newline
\indent
There are several conceptual differences between Poisson and multinomial sampling. Two reasons make the latter simpler for our purposes. First, Poisson sampling requires the introduction of an intensity parameter $\bar N$ on which the statistics of the map depend. A careful analysis requires then marginalization of this parameter. Second, the number of galaxies on each simulated map is the same for multinomial sampling. As a consequence the shot-noise term in the galaxy power spectrum is a constant. We do not need to consider it or subtract it in our analysis, as constant does not affect the information content of a statistic.
\newline
\indent
Naively one might be worried that the absence of fluctuations in the number of galaxies for multinomial sampling underestimates the total variance of the count maps. However, we found that this lower stochasticity precisely corresponds to the careful treatment of the shot noise term in the case of Poisson sampling.  In more details, we found that the following approach is equivalent to the multinomial sampling adopted in the text :  
\begin{enumerate}
\item After generation of the lognormal field, we use Poisson sampling to obtain the count map. To this effect, the intensity parameter of the poisson sampling in cell $i$ is taken to be
\beq
\bar N_i = f_i \rho_{g,i} \bar N 
\enq
where $\bar N$ is a free parameter, interpreted as the ensemble average number of galaxies in a totally unmasked cell. In the case of the CFHTLS data, we estimated it through
\beq \label{nbarest}
\bar N = \frac{N_{\textrm{tot}}}{\sum_i f_i}. 
\enq 
\item In the galaxy power spectrum we then subtract the shot-noise term. This requires first for each simulated map an estimate $\hat{\bar N} $ of $\bar N$. We use for this the above equation \eqref{nbarest}, where $N_{\textrm{tot}}$ now varies from map to map. Defining then $\delta_{g,i} = N /(f_i\hat {\bar N}) - 1 $ we can calculate the shot-noise contribution, that we then subtract to the spectrum. It is given by
\beq
P^{\textrm{shot}}_{g}(k) =  \frac{V}{d\hat{ \bar N}} \lp \frac {1}{d} \sum_{f_i \ne 0} \frac 1 {f_i}\rp.   = \frac 1 {\hat {\bar n}} \lp \frac 1 d \sum_{i} f_i \rp \lp \frac 1 d \sum_{f_i \ne 0} \frac 1 {f_i} \rp.
\enq
On the right hand side $\hat {\bar n}$ is the observed density of galaxies $N_{\textrm{tot}} /V $, and $d$ the number of cells.
Note that there is no need to subtract a shot-noise term in the $A^*$ power spectrum even for Poisson sampling, as the non-linear transformation already takes into the noise properties of the data.  
\end{enumerate}
Figure \ref{figure:PoissonVSmulti} illustrates that the two procedures leads to almost identical results.
Shown is the posterior for $w_0$ using $\delta_g$ or $A^{*}$ using the redshift bin $0.6 <z< 0.8$ of the W1 field of CFHTLS data, 
using Poisson sampling (dotted) as described above and using
multinomial sampling (solid) as described in the main text. The curves are virtually indistinguishable.

\end{appendix}

\bibliographystyle{mn2e}
\bibliography{Notes}

\begin{thebibliography}{}

\bibitem[\protect\citeauthoryear{{Carron} \& {Neyrinck}}{{Carron} \&
  {Neyrinck}}{2012}]{CarronNeyrinck2012}
{Carron} J.,  {Neyrinck} M.~C.,  2012, \apj, 750, 28

\bibitem[\protect\citeauthoryear{{Carron} \& {Szapudi}}{{Carron} \&
  {Szapudi}}{2013}]{Carronetal13}
{Carron} J.,  {Szapudi} I.,  2013, \mnras, 434, 2961

\bibitem[\protect\citeauthoryear{{Carron} \& {Szapudi}}{{Carron} \&
  {Szapudi}}{2014a}]{Carronetal14a}
{Carron} J.,  {Szapudi} I.,  2014a, \mnras, 439, L11

\bibitem[\protect\citeauthoryear{{Carron} \& {Szapudi}}{{Carron} \&
  {Szapudi}}{2014b}]{Carronetal14b}
{Carron} J.,  {Szapudi} I.,  2014b, ArXiv e-prints

\bibitem[\protect\citeauthoryear{{Carron}, {Wolk} \& {Szapudi}}{{Carron}
  et~al.}{2014}]{Carronetal14c}
{Carron} J.,  {Wolk} M.,    {Szapudi} I.,  2014, \mnras, 444, 994

\bibitem[\protect\citeauthoryear{{Cole} et~al.,}{{Cole}
  et~al.}{2005}]{Coleetal05}
{Cole} S.  et~al., 2005, \mnras, 362, 505

\bibitem[\protect\citeauthoryear{{Cooray} \& {Sheth}}{{Cooray} \&
  {Sheth}}{2002}]{Coorayetal02}
{Cooray} A.,  {Sheth} R.,  2002, Phys. Rep., 372, 1

\bibitem[\protect\citeauthoryear{{Coupon} et~al.,}{{Coupon}
  et~al.}{2012}]{Couponetal12}
{Coupon} J.  et~al., 2012, \aap, 542, A5

\bibitem[\protect\citeauthoryear{{de Putter}, {Wagner}, {Mena}, {Verde} \&
  {Percival}}{{de Putter} et~al.}{2012}]{dePutteretal12}
{de Putter} R.,  {Wagner} C.,  {Mena} O.,  {Verde} L.,    {Percival} W.~J.,
  2012, \jcap, 4, 19

\bibitem[\protect\citeauthoryear{{Joachimi} \& {Taylor}}{{Joachimi} \&
  {Taylor}}{2011}]{Joachimietal11}
{Joachimi} B.,  {Taylor} A.~N.,  2011, \mnras, 416, 1010

\bibitem[\protect\citeauthoryear{{Kravtsov}, {Berlind}, {Wechsler}, {Klypin},
  {Gottl{\"o}ber}, {Allgood} \& {Primack}}{{Kravtsov}
  et~al.}{2004}]{Kravtsovetal04}
{Kravtsov} A.~V.,  {Berlind} A.~A.,  {Wechsler} R.~H.,  {Klypin} A.~A.,
  {Gottl{\"o}ber} S.,  {Allgood} B.,    {Primack} J.~R.,  2004, \apj, 609, 35

\bibitem[\protect\citeauthoryear{{Li}, {Hu} \& {Takada}}{{Li}
  et~al.}{2014}]{LiEtal14}
{Li} Y.,  {Hu} W.,    {Takada} M.,  2014, \prd, 89, 083519

\bibitem[\protect\citeauthoryear{{Limber}}{{Limber}}{1954}]{Limber54}
{Limber} D.~N.,  1954, \apj, 119, 655

\bibitem[\protect\citeauthoryear{{Ma} \& {Fry}}{{Ma} \& {Fry}}{2000}]{Maetal00}
{Ma} C.-P.,  {Fry} J.~N.,  2000, \apj, 543, 503

\bibitem[\protect\citeauthoryear{{Navarro}, {Frenk} \& {White}}{{Navarro}
  et~al.}{1997}]{Navarroetal97}
{Navarro} J.~F.,  {Frenk} C.~S.,    {White} S.~D.~M.,  1997, \apj, 490, 493

\bibitem[\protect\citeauthoryear{{Neyrinck} \& {Szapudi}}{{Neyrinck} \&
  {Szapudi}}{2007}]{Neyrincketal07}
{Neyrinck} M.~C.,  {Szapudi} I.,  2007, \mnras, 375, L51

\bibitem[\protect\citeauthoryear{{Neyrinck}, {Szapudi} \& {Rimes}}{{Neyrinck}
  et~al.}{2006}]{Neyrincketal06}
{Neyrinck} M.~C.,  {Szapudi} I.,    {Rimes} C.~D.,  2006, \mnras, 370, L66

\bibitem[\protect\citeauthoryear{{Neyrinck}, {Szapudi} \& {Szalay}}{{Neyrinck}
  et~al.}{2009}]{Neyrincketal09}
{Neyrinck} M.~C.,  {Szapudi} I.,    {Szalay} A.~S.,  2009, \apjl, 698, L90

\bibitem[\protect\citeauthoryear{{Peacock} \& {Smith}}{{Peacock} \&
  {Smith}}{2000}]{Peacock00}
{Peacock} J.~A.,  {Smith} R.~E.,  2000, \mnras, 318, 1144

\bibitem[\protect\citeauthoryear{{Peebles}}{{Peebles}}{1980}]{Peebles1980}
{Peebles} P.~J.~E.,  1980, {The large-scale structure of the universe}

\bibitem[\protect\citeauthoryear{{Planck Collaboration} et~al.,}{{Planck
  Collaboration} et~al.}{2014}]{Planck14}
{Planck Collaboration} et~al., 2014, \aap, 571, A16

\bibitem[\protect\citeauthoryear{{Rimes} \& {Hamilton}}{{Rimes} \&
  {Hamilton}}{2005}]{Rimesetal05}
{Rimes} C.~D.,  {Hamilton} A.~J.~S.,  2005, \mnras, 360, L82

\bibitem[\protect\citeauthoryear{{Rimes} \& {Hamilton}}{{Rimes} \&
  {Hamilton}}{2006}]{Rimesetal06}
{Rimes} C.~D.,  {Hamilton} A.~J.~S.,  2006, \mnras, 371, 1205

\bibitem[\protect\citeauthoryear{{Scoccimarro}, {Sheth}, {Hui} \&
  {Jain}}{{Scoccimarro} et~al.}{2001}]{Scoccimarro01}
{Scoccimarro} R.,  {Sheth} R.~K.,  {Hui} L.,    {Jain} B.,  2001, ApJ, 546, 20

\bibitem[\protect\citeauthoryear{{Seljak}}{{Seljak}}{2000}]{Seljak00}
{Seljak} U.,  2000, \mnras, 318, 203

\bibitem[\protect\citeauthoryear{{Seo}, {Sato}, {Dodelson}, {Jain} \&
  {Takada}}{{Seo} et~al.}{2011}]{Seoetal11}
{Seo} H.-J.,  {Sato} M.,  {Dodelson} S.,  {Jain} B.,    {Takada} M.,  2011,
  \apjl, 729, L11

\bibitem[\protect\citeauthoryear{{Sheth} \& {Tormen}}{{Sheth} \&
  {Tormen}}{1999}]{Sethetal99}
{Sheth} R.~K.,  {Tormen} G.,  1999, \mnras, 308, 119

\bibitem[\protect\citeauthoryear{{Szapudi}}{{Szapudi}}{2009}]{Szapudi2009}
{Szapudi} I.,  2009, in {Mart{\'{\i}}nez} V.~J.,  {Saar} E.,
  {Mart{\'{\i}}nez-Gonz{\'a}lez} E.,   {Pons-Border{\'{\i}}a} M.-J.,  eds,
  Lecture Notes in Physics, Berlin Springer Verlag Vol. 665, Data Analysis in
  Cosmology. pp 457--492

\bibitem[\protect\citeauthoryear{{Szapudi} \& {Colombi}}{{Szapudi} \&
  {Colombi}}{1996}]{Szapudi96}
{Szapudi} I.,  {Colombi} S.,  1996, \apj, 470, 131

\bibitem[\protect\citeauthoryear{{Takada} \& {Hu}}{{Takada} \&
  {Hu}}{2013}]{Takadaetal13}
{Takada} M.,  {Hu} W.,  2013, \prd, 87, 123504

\bibitem[\protect\citeauthoryear{{Takada} \& {Jain}}{{Takada} \&
  {Jain}}{2009}]{Takadaetal09}
{Takada} M.,  {Jain} B.,  2009, \mnras, 395, 2065

\bibitem[\protect\citeauthoryear{{Tegmark} et~al.,}{{Tegmark}
  et~al.}{2004}]{Tegmarketal04}
{Tegmark} M.  et~al., 2004, \prd, 69, 103501

\bibitem[\protect\citeauthoryear{{Tinker}, {Weinberg}, {Zheng} \&
  {Zehavi}}{{Tinker} et~al.}{2005}]{Tinkeretal05}
{Tinker} J.~L.,  {Weinberg} D.~H.,  {Zheng} Z.,    {Zehavi} I.,  2005, \apj,
  631, 41

\bibitem[\protect\citeauthoryear{{Wolk}, {McCracken}, {Colombi}, {Fry},
  {Kilbinger}, {Hudelot}, {Mellier} \& {Ilbert}}{{Wolk} et~al.}{2013}]{Wolk13}
{Wolk} M.,  {McCracken} H.~J.,  {Colombi} S.,  {Fry} J.~N.,  {Kilbinger} M.,
  {Hudelot} P.,  {Mellier} Y.,    {Ilbert} O.,  2013, \mnras, 435, 2

\bibitem[\protect\citeauthoryear{{Zheng} et~al.,}{{Zheng}
  et~al.}{2005}]{Zhengetal05}
{Zheng} Z.  et~al., 2005, \apj, 633, 791

\bibitem[\protect\citeauthoryear{{Zheng}, {Coil} \& {Zehavi}}{{Zheng}
  et~al.}{2007}]{Zhengetal07}
{Zheng} Z.,  {Coil} A.~L.,    {Zehavi} I.,  2007, \apj, 667, 760

\end{thebibliography}

\end{document}